\documentclass[%
notitlepage,
superscriptaddress,
nofootinbib,
amsmath,
amssymb,
aps,
twocolumn
]{revtex4-1}

\usepackage{graphicx}
\usepackage{dcolumn}
\usepackage{bm}
\usepackage{mathtools}
\usepackage{amsmath}
\usepackage{amssymb}
\usepackage[hidelinks]{hyperref}
\usepackage{setspace}

\usepackage[dvipsnames, usenames]{xcolor}

\newcommand{\Kn}{\textnormal{Kn}}
\newcommand{\Ren}{\textnormal{Re}}

\begin{document}

\title{Causal, stable first-order viscous relativistic hydrodynamics with ideal gas microphysics}

\author{Alex Pandya}
\email{apandya@princeton.edu}
\affiliation{Department of Physics, Princeton University, Princeton, New Jersey
08544, USA}
\affiliation{Princeton Gravity Initiative, Princeton University, Princeton, NJ 08544, USA}
\author{Elias R. Most}
\email{emost@princeton.edu}
\affiliation{Princeton Gravity Initiative, Princeton University, Princeton, NJ 08544, USA}
\affiliation{Princeton Center for Theoretical Science, Princeton University, Princeton, NJ 08544, USA}
\affiliation{School of Natural Sciences, Institute for Advanced Study, Princeton, NJ 08540, USA}
\author{Frans Pretorius}
\email{fpretori@princeton.edu}
\affiliation{Department of Physics, Princeton University, Princeton, New Jersey
08544, USA}
\affiliation{Princeton Gravity Initiative, Princeton University, Princeton, NJ 08544, USA}

\date{\today}

\begin{abstract}
We present the first numerical analysis of causal, stable first-order relativistic hydrodynamics
with ideal gas microphysics, based in the formalism developed by
Bemfica, Disconzi, Noronha, and Kovtun (BDNK theory).  The BDNK approach
provides definitions for the conserved stress-energy tensor and baryon
current, and rigorously proves causality, local well-posedness, strong
hyperbolicity, and linear stability (about equilibrium) for the equations
of motion, subject to a set of coupled nonlinear inequalities involving the
undetermined model coefficients (the choice for which defines the
``hydrodynamic frame'').  We present a class of hydrodynamic frames derived
from the relativistic ideal gas ``gamma-law'' equation of state which
satisfy the BDNK constraints, and explore the properties of the resulting
model for a series of (0+1)D and (1+1)D tests in 4D Minkowski spacetime.
These tests include a comparison of the dissipation mechanisms in Eckart,
BDNK, and M\"uller-Israel-Stewart theories, as well as investigations of
the impact of hydrodynamic frame on the causality and stability properties
of Bjorken flow, planar shockwave, and heat flow solutions.
\end{abstract}

\maketitle

\section{Introduction} \label{sec:introduction}

Relativistic fluid theories have served as an essential tool in modeling a
remarkably diverse set of high-energy systems, examples of which range from
the quark-gluon plasma (QGP), a tiny soup of quarks and
gluons interacting
according to quantum chromodynamics \cite{Teaney:2009qa}, to the universe
itself, composed of $10^{11}$ galaxies \cite{Lauer2021}, dark matter, and
dark energy, all interacting gravitationally \cite{Andersson:2020phh}.  The
apparent differences between these two systems---and the many others well
described by fluid models---are bridged by the ubiquity of the two
requirements for applying hydrodynamics: (1) that interactions between
constituents occur often, and (2) that the interaction lengthscale $\ell$
is much smaller than the macroscopic size of the system, $L$
\cite{Rezzolla2013}.

The latter condition is commonly expressed in terms of a dimensionless
ratio known as the Knudsen number ${\Kn \equiv \ell/L}$, and hydrodynamics
is expected to be applicable when $\Kn \ll 1$.  In the limit $\Kn \to 0$,
interactions occur instantaneously, the system always remains in
thermodynamic equilibrium, and one observes a perfect (ideal) fluid. Real
fluids, however, have $\Kn \neq 0$ and pick up corrections beyond ideal
hydrodynamics.  At finite $\Kn$, equilibration takes finite time, and these
corrections drive the system toward thermodynamic equilibrium, for example
by allowing the transfer of energy (through heat conduction) or the
transfer of momentum (by viscosity).

Recent progress on the experimental front has made it possible to detect
these finite-$\Kn$ dissipative effects in relativistic systems, most
notable of which is the aforementioned QGP \cite{Romatschke:2017ejr},
though there are also significant indications that such effects may be
relevant to cosmological models \cite{Brevik:2014cxa,Brevik:2017msy} and
binary neutron star mergers \cite{Alford_2018,Most2021bulk,Hammond:2021vtv,Camelio:2022ljs,Most:2022yhe}.
This experimental progress has directed significant research interest
toward the theoretical development of relativistic dissipative
hydrodynamics, whose origins extend back as early as the 1940's with the
``relativistic Navier-Stokes'' theories of Eckart \cite{Eckart_1940} and
Landau-Lifshitz \cite{Landau_1987}.  These early models are built upon the
conserved currents of ideal hydrodynamics, and are likewise parameterized
by a set of ``hydrodynamic variables'' describing the thermodynamic state
of the system.  Dissipation is incorporated by adding to the conserved
currents first-order gradients of the hydrodynamic variables, which can be
roughly understood as corrections at linear order in $\Kn$.  Working at
finite $\Kn$ brings additional complications, however, in that the fluid
can now be outside of thermodynamic equilibrium, where the hydrodynamic
variables (e.g., temperature and flow velocity) no longer have a unique definition (the only restriction is that
they agree with thermodynamic observables upon equilibration).  Eckart and
Landau-Lifshitz each proposed their own method for fixing these definitions
(now referred to as fixing a ``hydrodynamic frame''), but it was later
shown that both prescriptions lead to fluid theories possessing acausal
solutions with unstable thermodynamic equilibria
\cite{Hiscock_1985,Hiscock_1983}, after which the theories were largely
abandoned in favor of a new approach developed by M\"uller, Israel, and
Stewart (MIS) \cite{Muller_1967,Israel:1976tn,Israel_1979}.  Their
formalism---henceforth MIS theory---involves promoting dissipative
corrections to independent degrees of freedom and evolving them in a way
that makes them \textit{gradually} approach their relativistic
Navier-Stokes values, rather than obtaining them instantaneously.  This
gradual application of dissipation appears to remedy the pathologies
present in the Eckart and Landau-Lifshitz theories, and MIS-type approaches
have since seen significant success in modeling heavy-ion collisions
\cite{Romatschke:2017ejr,PHENIX:2017djs}.

An alternative approach to fix the issues of the Eckart and Landau-Lifshitz
theories is to make a better choice for the hydrodynamic frame.  This is
precisely the method underlying the so-called BDNK formalism, developed by
and named for Bemfica, Disconzi, Noronha
\cite{Bemfica_2018,Bemfica:2020zjp} and Kovtun \cite{Kovtun_2019}.  
Different from MIS theories, these equations only include first-order
dissipative terms (i.e., bulk and shear viscosities, as well as heat
conductivity), leading to the same overall number of equations of motion as ideal
hydrodynamics, albeit now as second-order partial differential equations (PDEs). 
This leads to a simplification, compared to MIS theories which can capture
the full second-order dissipative sector, e.g. \cite{Denicol_2012}.
The major advantage of this restriction to first-order dissipation is the 
ability to \textit{rigorously prove} that the resulting theory is causal,
strongly hyperbolic, consistent with the second law of thermodynamics, and (linearly) stable
about thermodynamic equilibrium \cite{Bemfica:2020zjp}.  
Most of these properties have not been
proven for MIS-type theories, and examples of subtle pathologies have
arisen over the years\footnote{The original formulation of MIS theory has
  been shown to break down for sufficiently fast shockwaves
  \cite{Olson_1990_shocks}; many MIS simulations of heavy-ion collisions
  violate causality, unless a set of dynamical constraints are satisfied
  \cite{Bemfica:2020xym,Plumberg:2021bme}; and smooth solutions have been
  shown to break down in finite time (specifically in a formulation with
  only bulk viscosity) \cite{Disconzi:2020ijk}.}, obscured by the more
  complicated PDE structure.  
Finally, it has been shown that using a novel perturbative expansion of the Boltzmann equation allows one to derive BDNK hydrodynamics directly from kinetic theory \cite{Rocha:2022ind}, and that generalizing the entropy argument of Israel and Stewart \cite{Israel_1979} results in a variant of MIS theory containing BDNK as its first-order truncation \cite{Noronha:2021syv}.  These results underscore the ties between BDNK and MIS-type theories, and highlight that both approaches can be consistently derived from a description of the microscopic physics.

  These reasons motivate further investigation
  of BDNK theory as a potential alternative to MIS-based approaches.
  In part due to how recently the BDNK formalism was developed (the late
  2010's), all numerical studies up to this point have assumed a fluid with an
  underlying conformal symmetry and zero chemical potential (see, e.g.,
  \cite{Pandya_2021,Pandya:2022pif,Bantilan:2022ech}), a system often used
  as a toy model for a QGP in the limit of infinite temperature.  Most fluids of
  interest do not possess such a symmetry, however, and will generically
  have finite chemical potential.  Furthermore, the conformal model vastly
  reduces the theory's space of solutions, as the baryon current cannot
  backreact on the stress-energy tensor, the bulk viscosity vanishes
  identically, and there are no nontrivial heat flow solutions (in the
  sense that temperature gradients imply pressure gradients, so there are
  no dynamical solutions with a thermal gradient at constant pressure).

  In ideal hydrodynamics, generalizing from a conformal fluid to a
  non-conformal one would only require changing the equation of state.  In
  relativistic dissipative fluid theories the task is appreciably more
  difficult, however, as one must derive and constrain a set of so-called
  ``transport coefficients'' weighting the non-equilibrium corrections, bad
  choices for which lead to the pathologies that plagued the Eckart and
  Landau-Lifshitz theories.  In this work, we make use of the BDNK
  formalism to derive the first relativistic dissipative fluid model
  outside of the conformal limit which is causal, strongly hyperbolic,
  consistent with the second law of thermodynamics, and linearly stable
  about thermodynamic equilibrium.  Our model incorporates the microphysics
  of the relativistic ideal gas, and employs the ``gamma-law'' equation of
  state commonly used in astrophysics (see, e.g. \cite{Rezzolla2013}).  In
  this study, we define the model and derive all of the required transport
  coefficients, and then investigate the properties of the resulting theory
  in a set of test problems in 4D Minkowski spacetime with varying degrees
  of spatial symmetry.  Through these problems, we explore the new
  phenomena which arise outside of the conformal model, and also make use
  of the unique construction of BDNK theory---in particular, the fact that
  the theory provides allowed ranges for the transport coefficients---to
  study \textit{where} and \textit{why} pathologies arise when violating
  these constraints.  Such an investigation is motivated in light of the
  fact that the pathologies of the Eckart and Landau-Lifshitz theories went
  largely unnoticed for decades, and that even recent studies have shown
  that acausality (in the sense of possessing superluminal
  characteristics) appears ``asymptomatically'' in numerical simulations
  \cite{Plumberg:2021bme}, often going unnoticed.

In these tests, we investigate the dissipation mechanism of BDNK
theory, and show that in many ways it resembles the ``relaxation''
structure characteristic of MIS-type theories.  We find that the
qualitative behavior of solutions is typically unaffected by the
presence of ``slightly'' superluminal characteristics, and for problems
with sufficient spatial symmetry we observe no such change even as the
maximum local characteristic speed is taken to infinity.  Such cases do
come with numerical challenges, however, as the equations become
``stiff'' and require exceedingly small timesteps to integrate as the
characteristic speeds are increased.  For less symmetric problems, we
find that cases with sufficiently superluminal characteristics exhibit
a very fast instability, which may be related to the fact that
strictly subluminal characteristic speeds are required for the proof of
linear stability in \cite{Bemfica:2020zjp}.  One can also excite
instabilities for strictly causal solutions by violating the BDNK linear
stability constraints.  We consider such cases here as well, and again find
that small violations of the constraints appear without a qualitative
change in the solution; instabilities only set in for choices of parameters
well outside the provided bounds.

We begin in Sec. \ref{sec:model} with an overview of relativistic
hydrodynamics outside of equilibrium, and briefly review the BDNK
formalism.  We define the BDNK conserved currents, and summarize the
microphysics of interest for this work before providing a class of
hydrodynamic frames consistent with the proofs of \cite{Bemfica:2020zjp}.
In Sec. \ref{sec:eq_states}, we investigate the behavior of BDNK theory on
simple isotropic equilibrium states, and compare its dissipation mechanism
against that of Eckart theory and an example from the class of MIS
theories.  In Sec. \ref{sec:Bjorken_flow}, we analyze the effect of the
relaxation times on the qualitative behavior of Bjorken-type uniformly
expanding flows, and comment on the impact of superluminal
characteristics on the solution.  We extend this investigation with a
discussion of planar shockwave solutions in Sec. \ref{sec:shockwaves}, and
conclude our results in Sec. \ref{sec:heat_flow} with an investigation of
``pure'' heat flow solutions, which exist for the ideal gas model but not
for a conformal fluid, and an analysis of how hydrodynamic frame impacts
causality and stability of these solutions.  We briefly overview and
discuss our results in Sec. \ref{sec:conclusion}, and include a detailed
explanation of the choice of transport coefficients in Appendix
\ref{sec:hydro_frame}.  Details about the numerical method and convergence
tests are included in Appendix \ref{sec:numerics}.

\section{Model} \label{sec:model}

In this study we focus on the properties of BDNK theory, and specialize to
fluids in (3+1)D Minkowski spacetime.  We use the ``mostly plus'' metric
signature $(-+++)$, and employ the Einstein summation convention with
spacetime indices $\{a, b, c, d, e\}$; we restrict the usage of the letters
$\{i, j, k\}$ to denote spatial indices.

The remainder of this section overviews the gradient expansion approach to
hydrodynamics from which BDNK theory is derived, defines the BDNK conserved
currents, and then derives the required transport coefficients from the
microphysics of the relativistic ideal gas ``gamma-law'' equation of state.

The equations of relativistic hydrodynamics are written with varying
notation in the literature; here we most closely follow the notation of
\cite{Bemfica:2020zjp}.  Table \ref{table:notation} summarizes our notation
and common alternatives from the literature.

\begin{table}[h]
\begin{tabular}{l|c|c}
Quantity & ~This work~    & ~Lit. alternatives~ \\ \hline
energy density            & $\epsilon$ & $e$ \\
specific internal energy  & $e$        & $\epsilon$ \\
pressure                  & $P$        & $p$ \\
energy density + pressure & $\rho$     & $e+p$ \\
rest mass density         & $m n$      & $\rho$ \\
adiabatic index           & $\Gamma$   & $\gamma$ \\
entropy density           & $s$        & $\tilde{s}, \bar{s}$ \\
entropy per particle      & $\bar{s}$  & $s, \tilde{s}$ \\
\hline
\end{tabular}
\caption{Summary of notation used in this work, as well as alternative notation commonly used in the literature (such as, for example, \cite{Rezzolla2013}). \label{table:notation}}
\end{table}

\subsection{The gradient expansion}

Most relativistic fluid models are constructed by positing definitions for conserved currents, usually a stress-energy tensor $T^{ab}$ and a baryon current $J^{a}$, which are parameterized by a set of hydrodynamic variables.  These variables are drawn from the thermodynamic description of the system, and often include quantities such as the energy density $\epsilon$, baryon number density $n$, and (timelike) flow velocity $u^{a}$; other choices of parameters are possible as well, though, as one may use the laws of thermodynamics to exchange these quantities for others (for example, $\epsilon, n$ may be exchanged for the local temperature and baryon chemical potential, $T$ and $\mu$).  Given the conserved currents, one may evolve the state of the fluid forward through time using the corresponding conservation laws
\begin{align}
\nabla_{a} T^{ab} &= 0 \label{eq:Tab_cons_law} \\
\nabla_{a} J^{a}  &= 0. \label{eq:Ja_cons_law}
\end{align}

The simplest (and most commonly used) relativistic fluid model is that of ideal (perfect fluid) hydrodynamics, which assumes the system is always locally in thermodynamic equilibrium and has conserved currents
\begin{align}
T^{ab}_{0} &= \epsilon u^{a} u^{b} + P \Delta^{ab} \label{eq:Tab_0}\\
J^{a}_{0}  &= n u^{a}, \label{eq:Ja_0}
\end{align}
where the tensor
\begin{equation}
\Delta^{ab} \equiv g^{ab} + u^{a} u^{b}
\end{equation} 
projects onto the space orthogonal to the flow velocity $u^{a}$ and the isotropic pressure $P = P(\epsilon, n)$ defines the equation of state.  Inserting (\ref{eq:Tab_0}-\ref{eq:Ja_0}) into (\ref{eq:Tab_cons_law}-\ref{eq:Ja_cons_law}) yields a set of coupled nonlinear first-order PDEs known as the relativistic Euler equations.  Since ideal hydrodynamics assumes the fluid to always be in local thermodynamic equilibrium, entropy-generating dissipative processes such as viscosity and heat conduction are neglected; to study these phenomena, one needs to generalize beyond the perfect fluid.

A modern approach to construct a non-equilibrium fluid theory is to use the so-called gradient expansion, which assumes (1) that the system is near enough to equilibrium that it may be parameterized by the same set of hydrodynamic variables used to describe equilibrium fluids and (2) that gradients of these variables (which may be thought of as fluctuations about equilibrium) can be treated as small quantities.  Given these assumptions, the ``near-equilibrium'' conserved currents may be written
\begin{equation} \label{eq:gradient_exp}
\begin{aligned}
T^{ab} &= T^{ab}_{0} + \mathcal{O}(\nabla) + \mathcal{O}(\nabla^2) + ... \\
J^{a}  &= J^{a}_{0}  + \mathcal{O}(\nabla) + \mathcal{O}(\nabla^2) + ... \\
\end{aligned}
\end{equation}
where the $\mathcal{O}(\nabla)$ term includes a linear combination of all one-gradient terms constructed from the hydrodynamic variables, the $\mathcal{O}(\nabla^2)$ term includes products of all one-gradient terms as well as two-gradient terms, so on and so forth.  Since each successive term is assumed to be smaller than the one before it, one should be justified in truncating the series at some order in gradients\footnote{The expansion (\ref{eq:gradient_exp}) should not be interpreted to mean that higher-order gradient theories are necessarily superior to lower-order ones. In fact, there is evidence that issues arise in theories beyond first order in gradients---see \cite{DESCHEPPER19741,Kovtun:2011np}.}.  BDNK theory is constructed up to first order in gradients, and drops terms of $\mathcal{O}(\nabla^2)$ and higher; most formulations of MIS theory include second-order gradients (though they are not derived by explicit reference to a gradient expansion) \cite{Baier:2007ix,Muller_1967,Israel:1976tn,Israel_1979}, and a third-order theory has been constructed but to our knowledge has not been applied \cite{Grozdanov_2016,Muronga:2010zz}.

To construct a fluid theory up to a given order in gradients, it is conventional to first decompose the stress-energy tensor and baryon current with respect to the flow velocity $u^{a}$ via
\begin{align}
T^{a b} =\,\, &\mathcal{E} u^a u^b + \mathcal{P} \Delta^{ab} + \mathcal{Q}^a u^b + \mathcal{Q}^b u^a + \mathcal{T}^{ab} \label{eq:Tab} \\
J^{a} =\,\, &\mathcal{N} u^{a} + \mathcal{J}^{a}, \label{eq:Ja}
\end{align}
where the script quantities are projections of the conserved currents defined by
\begin{equation} \label{eq:projections}
\begin{aligned}
&\mathcal{E} = u_{a} u_{b} T^{ab}, ~~~ \mathcal{P} = \frac{1}{3} \Delta_{a b} T^{ab}, ~~~ \mathcal{Q}^{a} = - \Delta^{a b} u^{c} T_{b c} \\
&\mathcal{T}^{ab} = T^{<ab>}, ~~~ \mathcal{N} = - u_{a} J^{a}, ~~~ \mathcal{J}^{a} = \Delta^{a b} J_{b}
\end{aligned}
\end{equation}
where the angle brackets around a pair of indices denote the traceless part of a rank-two tensor which is also orthogonal to the flow velocity in both indices, namely
\begin{multline}
X^{<ab>} \equiv \frac{1}{2} \Big( \Delta^{a c} \Delta^{b d} X_{cd} + \Delta^{a d} \Delta^{b c} X_{cd} \\
- \frac{2}{3} \Delta^{a b} \Delta^{c d} X_{cd} \Big).
\end{multline}
Inserting (\ref{eq:projections}) into (\ref{eq:Tab}-\ref{eq:Ja}) simply yields the identity; constructing a fluid model involves replacing (\ref{eq:projections}) with a set of constitutive relations defining the script quantities $\mathcal{E}, \mathcal{P}, \mathcal{Q}^{a}, \mathcal{T}^{ab}, \mathcal{N}, \mathcal{J}^{a}$ in terms of the hydrodynamic variables (here $\epsilon, n, u^{a}$).  For ideal hydrodynamics, ${\mathcal{E}_{0} = \epsilon, \mathcal{P}_{0} = P, \mathcal{N}_{0} = n}$, and the purely dissipative terms all vanish, ${\mathcal{Q}_{0}^{a} = \mathcal{T}_{0}^{ab} = \mathcal{J}_{0}^{a} = 0}$.

\subsection{BDNK theory}

The key result of the BDNK formalism is that a sensible relativistic viscous fluid theory may be constructed by adding terms of $\mathcal{O}(\nabla)$ to the conserved currents $T^{ab}, J^{a}$, contrary to the expectation that arose after the Eckart and Landau-Lifshitz theories were shown to be acausal and thermodynamically unstable in the 1980's \cite{Hiscock_1983,Hiscock_1985}.  The authors of \cite{Bemfica:2020zjp} were able to prove that a generalized version of Eckart's theory can be rendered causal, (linearly) stable about equilibrium, consistent with the second law of thermodynamics, and strongly hyperbolic, provided one makes good choices for the (transport) coefficients weighting the gradient terms.  In this work, we will use this generalized Eckart theory and derive a set of transport coefficients based in ideal gas microphysics which are consistent with the requirements laid out in \cite{Bemfica:2020zjp}, thus obtaining a theory with all of the aforementioned properties.

The conserved currents used in \cite{Bemfica:2020zjp} (hereafter referred to as the BDNK conserved currents) are defined to be (\ref{eq:Tab}-\ref{eq:Ja}) with the definitions
\begin{align}
&\mathcal{E} = \epsilon + \tau_{\epsilon} \big[ u^c \nabla_c \epsilon + \rho \nabla_c u^c \big] \label{eq:script_E} \\
&\mathcal{P} = P - \zeta \nabla_c u^c + \tau_{P} \big[ u^c \nabla_c \epsilon + \rho \nabla_c u^c \big] \label{eq:script_P} \\
&\mathcal{Q}^{a} = \tau_{Q} \rho u^c \nabla_c u^a + \beta_{\epsilon} \Delta^{a c} \nabla_c \epsilon + \beta_{n} \Delta^{a c} \nabla_c n \label{eq:Q_a}\\
&\mathcal{T}^{ab} = - 2 \eta \sigma^{ab} \equiv -2 \eta \nabla^{<a} u^{b>} \label{eq:script_T_ab} \\
&\mathcal{N} = n \label{eq:script_N}\\
&\mathcal{J}^{a} = 0. \label{eq:script_J_a}
\end{align}
Throughout we have used the shorthand
\begin{equation} \label{eq:rho}
\rho \equiv \epsilon + P.
\end{equation}
Note that these definitions imply the BDNK baryon current (\ref{eq:Ja}, \ref{eq:script_N}-\ref{eq:script_J_a}) is identical to that of ideal hydrodynamics (\ref{eq:Ja_0}).  The same is not true for the stress-energy tensor, however, as the terms $\mathcal{E}, \mathcal{P}$ now incorporate gradient corrections to the energy density and pressure, respectively; $\mathcal{Q}^{a}$ is the heat flux vector; and $\sigma^{ab}$ is known as the shear tensor.  The quantities $\beta_{i}$ appearing in $\mathcal{Q}^{a}$ are defined in terms of the substance's underlying microphysics by 
\begin{align}
\beta_{\epsilon} &= \tau_{Q} p'_{\epsilon} + \frac{\sigma}{\rho} \kappa_{\epsilon} \label{eq:beta_eps} \\
\beta_{n} &= \tau_{Q} p'_{n} + \frac{\sigma}{n} \kappa_{n}, \label{eq:beta_n}
\end{align}
where we have defined the shorthand
\begin{align}
p'_{\epsilon} &\equiv \Big( \frac{\partial P}{\partial \epsilon} \Big)_{n} \label{eq:pPeps_defn}\\
p'_{n} &\equiv \Big( \frac{\partial P}{\partial n} \Big)_{\epsilon} \\
\kappa_{\epsilon} &\equiv \frac{\rho^2 T}{n} \Big( \frac{\partial (\mu/T)}{\partial \epsilon} \Big)_{n} \\
\kappa_{n} &\equiv \rho T\Big( \frac{\partial (\mu/T)}{\partial n} \Big)_{\epsilon} \label{eq:kappa_n_defn} \\
\kappa_{s} &\equiv \kappa_{\epsilon} + \kappa_{n} \label{eq:kappa_s}
\end{align}
where $T(\epsilon, n)$ is the temperature of the fluid, $\mu(\epsilon, n)$ is the (relativistic) chemical potential, and subscripts on the parentheses mean that the subscripted quantity is being held constant while the derivative in parentheses is evaluated.

In (3+1)D spacetime, the system (\ref{eq:Tab_cons_law}-\ref{eq:Ja_cons_law}) 
provides five equations of motion to evolve the five unknowns $\epsilon, n$, and three components of $u^{a}$ (the fourth component is determined by the fact that $u^a$ is timelike, $u_{c} u^{c} = -1$).  To solve these equations, however, one must specify the equation of state $P(\epsilon, n)$ as well as the 8 transport coefficients, $\tau_{\epsilon}, \tau_{P}, \tau_{Q}, \eta, \zeta, \sigma, \beta_{\epsilon}, \beta_{n}$.  The latter naturally fall into three categories.  The first three are relaxation times $\tau_{\epsilon}, \tau_{P}, \tau_{Q}$, and they determine the timescales over which dissipation impacts the solution.  The coefficients $\eta, \zeta, \sigma$ determine the strength of the dissipative effects, and are known as the shear viscosity, bulk viscosity, and thermal conductivity, respectively.  Finally the coefficients $\beta_{\epsilon}, \beta_{n}$ appear only in $\mathcal{Q}^{a}$ and control the contributions of energy density and baryon density gradients to the heat flux.  The choice for these 8 coefficients defines the ``hydrodynamic frame'', and must satisfy the constraints laid out in \cite{Bemfica:2020zjp} in order to produce a hydrodynamic theory which is causal, strongly hyperbolic, and well-posed, with stable equilibrium states.

\subsection{Relativistic ideal gas microphysics}

In this study we specialize to fluids with the so-called gamma-law equation of state, derived from the thermodynamics of the relativistic ideal gas:
\begin{equation} \label{eq:EOS}
P(\epsilon, n) = [\Gamma - 1] m n \, e(\epsilon, n) = n \, T(\epsilon, n),
\end{equation}
where $\Gamma \in (1,2)$ is the adiabatic index.  The specific internal energy is defined to be the system's total internal energy divided by its total mass, $e \equiv U/(m N)$ (where $N = n V$ is the total number of particles, $V$ is the system volume), and is related to the total energy density by
\begin{equation} \label{eq:e_defn}
\epsilon = m n (1 + e).
\end{equation}
One can see that (\ref{eq:e_defn}) is simply the statement that the total energy density is the sum of the rest mass energy density $m n$ and the internal energy density $m n e = U/V$.

Note that (\ref{eq:beta_eps}-\ref{eq:beta_n}) also require us to take derivatives of the quantity $\frac{\mu}{T}(\epsilon, n)$.  This quantity can be computed using the laws of thermodynamics, specifically the Euler relation ${U = T S - P V + \mu_{N} N}$ (where $\mu_{N}$ is the Newtonian chemical potential)
which may be written using (\ref{eq:e_defn}) as
\begin{equation} \label{eq:Euler_relation}
\rho = T s + n \mu,
\end{equation}
where $\mu \equiv \mu_{N} + m$ is the relativistic chemical potential.  We will need to compute the entropy density $s \equiv S/V$, which can be done using the first law of thermodynamics ${dU = T dS - P dV + \mu dN}$ written in the form 
\begin{equation}
de = T d \Big( \frac{s}{n m} \Big) - P d \Big( \frac{1}{n m} \Big),
\end{equation}
which we can expand, divide by $d n$, substitute for $P, T$ using (\ref{eq:EOS}), and integrate to find
\begin{equation} \label{eq:S_over_V}
s(\epsilon, n) = m n \bigg( \frac{1}{(\Gamma - 1) m} \ln \bigg[ \frac{e(\epsilon, n)}{n^{\Gamma - 1}} \bigg] + \textnormal{const} \bigg).
\end{equation}
Inserting (\ref{eq:S_over_V}) into (\ref{eq:Euler_relation}) yields the final result
\begin{equation} \label{eq:mu}
\mu(\epsilon, n) = m + m e(\epsilon, n) \bigg( \Gamma - \ln \bigg[ \frac{e(\epsilon, n)}{n^{\Gamma - 1}}  \bigg] + \textnormal{const} \bigg).
\end{equation}

Using (\ref{eq:EOS}-\ref{eq:e_defn},\ref{eq:mu}) one can now compute all of the required microphysics-based derivatives (\ref{eq:pPeps_defn}-\ref{eq:kappa_n_defn}), which are:
\begin{align}
p'_{\epsilon} &= \Gamma - 1 \\
p'_{n} &= -(\Gamma - 1) m \\
\kappa_{\epsilon} &= -(\Gamma - 1) \frac{\epsilon \rho^2}{n^2 P} \\
\kappa_{n} &= \frac{\rho}{n^2 P} \big[ (\Gamma - 1) \epsilon^2 + P^2 \big]
\end{align}
so
\begin{align}
\beta_{\epsilon} &= (\Gamma - 1) \tau_{Q} - (\Gamma - 1) \frac{\sigma \epsilon \rho}{n^2 P}\\
\beta_{n} &= -(\Gamma - 1) m \tau_{Q} + \frac{\sigma \rho}{n^3 P} \big[ (\Gamma - 1) \epsilon^2 + P^2 \big] 
\end{align}
for the relativistic ideal gas.  

We will also need to compute a few other quantities from the microphysics, including the sound speed (squared)
\begin{equation} \label{eq:cs_sq}
c_s^2 \equiv \Big( \frac{\partial P}{\partial \epsilon} \Big)_{\bar{s}} = \Big( \frac{\partial P}{\partial \epsilon} \Big)_{n} + \frac{n}{\rho} \Big( \frac{\partial P}{\partial n} \Big)_{\epsilon} = \frac{\Gamma P}{\rho}
\end{equation}
where $\bar{s} \equiv S/N$ is the entropy per particle, and
\begin{align}
\kappa_{s} &= -(\Gamma - 1) m \frac{\rho}{n} \\
\omega &\equiv \frac{\kappa_{s}}{\kappa_{\epsilon}} = \frac{m n P}{\epsilon \rho} \label{eq:omega} \\
\alpha &\equiv \frac{p'_{\epsilon}}{c_s^2} = \frac{\Gamma - 1}{c_s^2} \label{eq:alpha}
\end{align}
which appear in our choice of hydrodynamic frame in the following subsection as well as the linear stability constraints (see Appendix \ref{sec:hydro_frame}).

\subsection{Hydrodynamic frame}

In the following, we will discuss our choice of hydrodynamic frame,
which can critically affect the outcome of the evolution, i.e., only
certain choices of hydrodynamic frames will be causal, stable about equilibrium states, consistent with the second law of thermodynamics, and strongly hyperbolic.
We will provide a more in-depth discussion of the implications of 
this frame choice in Sec. \ref{sec:eq_states}, and limit ourselves to
just specifying our choice of frame in this subsection.

The evolution of the dissipative sector is governed by six transport
coefficients: the
relaxation times $\tau_{\epsilon}, \tau_{P}, \tau_{Q}$ and the ``physical''
dissipative coefficients $\eta, \zeta, \sigma$,
all of which will typically depend on the hydrodynamic variables
(here $\epsilon, n$).
For these six quantities,
we introduce a new class of hydrodynamic frames defined by
\begin{equation} \label{eq:hydro_frame}
\begin{aligned}
&\eta = \rho c_s^2 L \, \hat{\eta}, ~~~ \zeta = \rho c_s^2 L \, \hat{\zeta}, ~~~ \sigma = \frac{\hat{V} L \rho c_s^2}{(-\kappa_{\epsilon})} \, \hat{\sigma} \\
&\tau_{\epsilon} = \tau_{Q} = L \hat{V} \, \hat{\tau} , ~~~ \tau_{P} = 2 (\Gamma - 1) L \hat{V} \\
\end{aligned}
\end{equation}
where all of the hatted quantities are dimensionless, and we have defined
the shorthand
\begin{align}
V &\equiv \frac{4 \eta}{3} + \zeta \label{eq:V} \\
\hat{V} &\equiv \frac{V}{\rho c_s^2 L} \equiv \Ren^{-1}, \label{eq:Vhat_defn}
\end{align}
so $V$ is a combined viscosity.  The quantity $\hat{V}$ acts like an
inverse Reynolds number\footnote{The nonrelativistic Reynolds number is
  typically defined to be $\Ren_{NR} \equiv (m n) v L/\eta$, where $v$ is
  the flow speed and $L$ is a characteristic length.  Our effective inverse
  Reynolds number (\ref{eq:Vhat_defn}) differs in that the mass density is
replaced by a measure of the total energy density $\rho$, $v$ is replaced
by $c_s^2$ (which must be divided by a factor of $c$ in units where $c \neq
1$) and $\eta \to V$.} in the sense that it is a dimensionless ratio
involving the viscosity $V$, a measure of the (energy) density $\rho$, a
characteristic flow speed $c_s$, and a characteristic lengthscale $L > 0$.
Note that we have factored the transport coefficients so that they are proportional to a dimensionless free parameter (denoted with a hat) and a dimensionful free parameter (the lengthscale $L$).  The dimensionless parameters $\hat{\eta}, \hat{\zeta}, \hat{\sigma}, \hat{\tau}$ may be freely modulated according to the physics one wants to investigate.  The dimensionful parameter, however, should be set based upon the lengthscale of interest, for example by specifying the total viscosity $V$, an effective Reynolds number $\Ren$, then solving (\ref{eq:Vhat_defn}) for $L$; another option would be to set $L \sim \tau_{i}/v$, where $\tau_{i}$ is a relaxation time and $v$ is the three-velocity of the flow; see also, for example, the power-counting scheme of \cite{Denicol:2012cn}.
For the sake of simplicity, we set $L = 1$ in this work.

In this class of hydrodynamic frames, the quantities $\hat{\eta} > 0,
\hat{\zeta} \geq 0, \hat{\sigma} \geq 0$ may be treated as free parameters
determining the amount of shear viscosity, bulk viscosity, and thermal
conductivity in the model.  The three relaxation times have only a
single free parameter $\hat{\tau}$, which 
fixes the characteristic speeds. It is important to point out that for
this reason the class of hydrodynamic frames (\ref{eq:hydro_frame}) is
not the most general one.  We find that this restriction is not too
severe, however, as we are only ever interested in fixing a single
characteristic speed, namely the maximum speed $c_+$ (\ref{eq:cpmsq}), as
it determines causality and the stability of fast flows (see Sec.
\ref{sec:shockwaves}).  The benefit of this restriction is made clear
when attempting to satisfy the BDNK constraints, which can be done
analytically in this case; see Appendix \ref{sec:hydro_frame}.

The set of frame constraints laid out in \cite{Bemfica:2020zjp} for the frame (\ref{eq:hydro_frame}-\ref{eq:Vhat_defn}) are all satisfied provided
\begin{equation} \label{eq:simple_constraints}
\hat{\sigma} \leq \frac{1}{3}, ~~~ \hat{\tau} \geq \frac{(\Gamma - 1)(2 - c_s^2) + c_s^2}{1 - c_s^2},
\end{equation}
where the constraint on $\hat{\sigma}$ ensures linear stability and the
$\hat{\tau}$ constraint ensures causality. The causality constraint in
(\ref{eq:simple_constraints}) is simpler\footnote{An even simpler 
  causal bound on $\hat{\tau}$ may be obtained by taking $\Gamma
  \to 2$, which yields $\hat{\tau} \geq \frac{2}{1 - c_s^2}$ and bounds $|c_+|$ to be smaller than the causality constraint in (\ref{eq:simple_constraints}).} 
and places a higher lower bound on $\hat{\tau}$ than the precise BDNK causality constraints derived from the frame ansatz (\ref{eq:hydro_frame}); for those, see Appendix \ref{sec:hydro_frame}.
Note that $c_s^2 \to 1$ requires $\hat{\tau} \to \infty$ in order for the characteristic speeds to remain subluminal; this is a limitation of the frame ansatz (\ref{eq:hydro_frame}), and for systems in which the sound speed is expected to be very close to the speed of light one would likely need a different hydrodynamic frame.

\section{Results} \label{sec:results}
In this section we explore the properties of the model presented in Sec.
\ref{sec:model}.  In particular, we investigate the behavior of solutions
to the equations in (3+1)D Minkowski spacetime, assuming various degrees of
spatial symmetry. We focus on dissipative effects missing in
the previous studies \cite{Pandya_2021,Pandya:2022pif} which considered a
conformal fluid at zero baryon chemical potential,
namely bulk viscosity and thermal conductivity.
That said, we only consider test problems in flat spacetime with variation
in at most one spatial dimension; this high degree of symmetry renders the
bulk and shear viscosities degenerate in the sense that they only appear in
the combination $V$ (\ref{eq:V}) in the equations of motion. 
We can still study the effect of this combined viscosity here, however, and
the spatial symmetry does not constrain the thermal conductivity (which we
consider in Sec. \ref{sec:heat_flow}).

In the subsections that follow, we explore how static (trivial) equilibrium
states illustrate that the BDNK equations incorporate similar
``relaxation''-type behavior to MIS-type theories (Sec.
\ref{sec:eq_states}), after which we consider how the choice of
hydrodynamic frame impacts uniformly expanding (Bjorken) flows (Sec.
\ref{sec:Bjorken_flow}).  We then generalize the analysis on steady-state
shockwave solutions from \cite{Pandya_2021} to arbitrary BDNK fluids, and
study solutions for the ideal gas microphysics presented earlier in Sec.
\ref{sec:shockwaves}.  We conclude the section by considering how the
thermal conductivity expands the space of nontrivial solutions, and further
consider how hydrodynamic frame determines the stability of such states in
Sec. \ref{sec:heat_flow}.

In order to obtain the numerical results presented throughout this section,
one must specify several model parameters.  Rather than repeatedly stating
these in the text, we summarize the parameters used to produce each of the
figures below in Table \ref{table:parameters}.

\begin{table}[h]
\renewcommand{\arraystretch}{1.2}
\begin{tabular}{c || c | c | c | c | c}
Figure                      & ~$\Gamma$~ & ~$m$~ & ~$\hat{V}$~ & ~$\hat{\sigma}$~ & $\hat{\tau}$ \\ \hline \hline
\ref{fig:bjorken}           & $\frac{4}{3}$ & $1$   & $\frac{1}{10}$ & $0$ & $0.5, 1, 2$ \\ 
\ref{fig:shockwave_profile} & $\frac{4}{3}$ & $0.1$ & $\frac{2}{15}$ & $0$ & $1.5$ \\
\ref{fig:shock_instability} & $\frac{4}{3}$ & $0.1$ & $\frac{4}{3}$ & $0$ & $1.5, 3$ \\
\ref{fig:acaus_instab}      & $\frac{4}{3}$ & $0.1$ & $\frac{4}{3}$ & $0$ & $0.25, 0.4, 0.5, 1.5$ \\
\ref{fig:heat_stationary}   & $\frac{4}{3}$ & $0.1$ & $\frac{2}{15}$ & $0, \frac{1}{3}$ & $1.5$ \\
\ref{fig:telegraphers}      & $\frac{4}{3}$ & $0.1$ & $\frac{2}{15}$ & $0.15, 1.5, 7.5$ & $1.5, 15, 75$ \\
\end{tabular}
\caption{Parameters used in the numerical tests whose results are shown in Figs. \ref{fig:bjorken}-\ref{fig:telegraphers}.} \label{table:parameters}
\end{table}

\subsection{Trivial equilibrium states and the structure of BDNK theory} \label{sec:eq_states}

In this section, we aim to briefly compare the structure of Eckart, MIS,
and BDNK theories to better understand the cause of pathologies (in the
case of Eckart theory) and how those pathologies are avoided (in MIS and
BDNK theories). 
Eckart theory can be obtained from BDNK theory in a very
simple way, by taking the BDNK conserved currents
(\ref{eq:Tab}-\ref{eq:Ja}), and setting 
\begin{equation} \label{eq:Eckart_frame}
\textnormal{Eckart frame: } \tau_{\epsilon} = \tau_{P} = 0, ~~~ \tau_{Q} = -\frac{\kappa T}{\rho},
\end{equation}
where\footnote{The Eckart heat flux vector is typically written
  ${\mathcal{Q}^{a}_{E} = -\kappa T(u^{c}\nabla_{c} u^{a} + \Delta^{a c}/T
  \, \nabla_{c} T)}$, which can be obtained from
  (\ref{eq:Q_a},\ref{eq:Eckart_frame}-\ref{eq:kappa}) and the thermodynamic
  identity (\ref{eq:thermo_identity}).  Written in this form, the BDNK heat
  flux $\mathcal{Q}^{a}$ is simply the Eckart heat flux
  $\mathcal{Q}^{a}_{E}$ plus the transverse projection of the relativistic
  Euler equations (\ref{eq:vector_reg_term}), which is second order in
  gradients on-shell. \label{footnote:Eckart_heat_flux}} the thermal
  conductivity coefficient is
\begin{equation} \label{eq:kappa}
\kappa \equiv \frac{\sigma \rho^2}{n^2 T}
\end{equation}
and the Eckart particle current is given by $J^{a}_{0}$ (\ref{eq:Ja_0}).
Note that Eckart theory badly violates the assumptions underlying the BDNK
formalism's proofs of causality, linear stability, and strong
hyperbolicity, each of which takes $\tau_{\epsilon}, \tau_{P}, \tau_{Q} >
0$.

MIS-type theories can be written in the form
\begin{equation}
T^{ab}_{MIS} = T^{ab}_{0} + \pi^{ab}
\end{equation}
where $T^{ab}_{0}$ is the perfect fluid stress-energy tensor
(\ref{eq:Tab_0}) and we will for the sake of comparison take the MIS baryon
current to be (\ref{eq:Ja_0}).  The tensor $\pi^{ab}$ incorporates all of
the dissipative effects present in the theory, and is evolved using a set
of ``relaxation-type'' equations of the form
\begin{equation} \label{eq:MIS_relax}
u^{c} \nabla_{c} \pi^{ab} = \frac{1}{\tau_{\pi}} \big( \pi^{ab}_{NS} - \pi^{ab} \big) + I^{ab},
\end{equation}
where the tensor $\pi^{ab}_{NS}$ includes dissipative effects from a
relativistic Navier-Stokes theory (typically either Eckart theory or Landau-Lifshitz theory elsewhere), and $I^{ab}$ are
additional higher-order corrections not present in $\pi^{ab}_{NS}$.  For the
sake of simplicity we will take $I^{ab} = 0$. 
We define $\pi^{ab}_{NS}$ to be the most general linear combination of terms with a single gradient of the hydrodynamic variables (i.e. the general-frame first-order dissipative correction)
\begin{equation}
\begin{aligned}
\pi^{ab}_{NS} &= \mathcal{A} u^{a} u^{b} + \mathcal{B} \Delta^{ab} + \mathcal{C}^{a} u^{b} + \mathcal{C}^{b} u^{a} + \mathcal{D}^{ab} \\
\mathcal{A} &= c_1 u^{c} \nabla_{c} \epsilon + c_2 u^{c} \nabla_{c} n + c_3 \nabla_{c} u^{c},
\end{aligned}
\end{equation}
where we have performed the same decomposition on $\pi^{ab}$ as was done for $T^{ab}$ to get (\ref{eq:Tab}), so $u_{a} \mathcal{C}^{a} = u_{a} \mathcal{D}^{ab} = 0$.  We will only need the definition of $\mathcal{A}$ in the analysis below, and we weight the three unique $\mathcal{O}(\nabla)$ scalar terms with undetermined gradient-free coefficients $c_1, c_2, c_3$.  Note that because we are taking the most general definition for $\pi^{ab}_{NS}$ up to $\mathcal{O}(\nabla)$, $u_{a} u_{b} \pi^{ab}_{NS} \neq 0$, different from most implementations of MIS theory (which use the Landau frame wherein $\mathcal{A} = \mathcal{C}^{a} = 0$).

To compare the structure of the Eckart, MIS, and BDNK theories, we will
consider a very restrictive set of initial data,
\begin{equation} \label{eq:eq_state_ID}
\begin{aligned}
&\epsilon, n \neq 0, &&\epsilon_{,i} = n_{,i} = u^{i} = 0, &&&\textnormal{(Eckart, BDNK, MIS)} \\
&\dot{u}^{i} = 0, &&\, &&&\textnormal{(Eckart)} \\
&\dot{\epsilon} \neq 0, &&\dot{u}^{i} = 0, &&&\textnormal{(BDNK)} \\
&\pi^{tt} \neq 0, &&\pi^{ab}_{,i} = \pi^{ai} = 0, &&&\textnormal{(MIS)}
\end{aligned}
\end{equation}
where $\dot{X} \equiv \partial_{t} X$, $X_{,i} \equiv \partial_{i} X$, $a$
is a spacetime index, $i$ is a purely spatial index, and all quantities
above are independent of $x^{i}$ making the system spatially isotropic.
Note that in general the Eckart and BDNK theories include second-order PDEs in
$\epsilon, u^{a}$, and thus we must specify time derivatives
$\dot{\epsilon}, \dot{u}^{i}$ at $t = 0$ (though in Eckart theory $\dot{\epsilon}$ only appears in the heat flux vector, which vanishes due to spatial symmetry here, leaving only $\dot{u}^{i}$ freely specifiable in (\ref{eq:eq_state_ID})).  MIS-type theories, on the other
hand, do not require specification of $\dot{\epsilon}, \dot{u}^{i}$, but do
require values for $\pi^{ab}$. To choose a set of data roughly consistent
between the Eckart, BDNK, and MIS theories, we set nearly all time
derivatives and components of $\pi^{ab}$ to zero, leaving only
$\dot{\epsilon}, \pi^{tt} \neq 0$ so that the ``energy density'' $T^{tt}$
has a ``non-equilibrium'' contribution, if the theory allows for one
(Eckart theory does not). 

Starting with the baryon current conservation law (\ref{eq:Ja_cons_law}),
we find that the initial data (\ref{eq:eq_state_ID}) for all three theories
implies 
\begin{equation}
\dot{n} = 0,
\end{equation}
so $n$ is a constant in space and in time.  Moving on to the stress-energy conservation law (\ref{eq:Tab_cons_law}), one finds only the $t$-component is nontrivial:
\begin{equation} \label{eq:eq_state_t_general}
T^{tt}_{,t} = 0 \implies T^{tt} = \textnormal{const},
\end{equation}
because all off-diagonal components and spatial derivatives in the conservation law vanish ($T^{ti} = T^{ij}_{,i} = 0$) due to the spatial isotropy of the system.  For Eckart theory, (\ref{eq:eq_state_t_general}) combined with the definition of the Eckart stress-energy tensor (\ref{eq:Tab},\ref{eq:script_E},\ref{eq:Eckart_frame}) implies
\begin{equation}
\epsilon = T^{tt}, ~~~ \textnormal{(Eckart)}
\end{equation}
and there are no dynamics.
For BDNK theory, (\ref{eq:eq_state_t_general}) instead entails
\begin{equation}
\epsilon + \tau_{\epsilon} \dot{\epsilon} = T^{tt}, ~~~ \textnormal{(BDNK)}
\end{equation}
which is a first-order ordinary differential equation (ODE) in $t$ and does have dynamics if $\dot{\epsilon}$ is chosen to be nonzero at $t = 0$.  Finally, MIS theory has equations of motion coming from (\ref{eq:eq_state_t_general}) as well as (\ref{eq:MIS_relax}):
\begin{equation} \label{eq:MIS_EOM}
\epsilon + \pi^{tt} = T^{tt}, ~~~ \dot{\pi}^{tt} = \frac{1}{\tau_{\pi}}(c_1 \dot{\epsilon} - \pi^{tt}), ~~~ \textnormal{(MIS)}
\end{equation}
after noting that 
$\Delta^{tt} = \mathcal{C}^{t} = \mathcal{D}^{tt} = 0$ because they are all orthogonal to $u^{a}$ by definition, and that $\nabla_{c} u^{c}$ vanishes due to spatial isotropy.
The left equation of (\ref{eq:MIS_EOM}) implies $\pi^{tt} = T^{tt} - \epsilon$, and its time derivative implies $\dot{\epsilon} = -\dot{\pi}^{tt}$.  Using these two facts in the equation of (\ref{eq:MIS_EOM}) on the right, we can eliminate $\pi^{tt}$ to obtain a single equation for MIS theory which can be directly compared to the Eckart and BDNK theory equations:
\begin{equation} \label{eq:theory_comp}
\begin{aligned}
\epsilon &= T^{tt}, &&\textnormal{(Eckart)} \\
\dot{\epsilon} &= \frac{1}{\tau_{\epsilon}} (T^{tt} - \epsilon), &&\textnormal{(BDNK)} \\
\dot{\epsilon} &= \frac{1}{\tau_{\pi} + c_1} (T^{tt} - \epsilon), &&\textnormal{(MIS)}
\end{aligned}
\end{equation}
and one can immediately see that the BDNK and MIS equations of motion are equivalent upon the identification $\tau_{\epsilon} = \tau_{\pi} + c_1 \equiv \tau$.

To understand (\ref{eq:theory_comp}), it is important to ask why the
BDNK/MIS theories possess dynamical solutions at all, since the system
being described is a ``trivial'' equilibrium state and has no dynamics in
the physical observables $T^{ab}, J^{a}$.  
We can most easily clarify this by considering 
the behavior of the hydrodynamic variables outside of equilibrium.  In particular, it is useful to begin by noting that the BDNK stress-energy tensor is composed of an equilibrium piece and a non-equilibrium one, i.e.
\begin{align}
u_\mu u_\nu T^{\mu\nu} = T^{tt} = \epsilon + \delta \epsilon.
\label{eq:general_frame}
\end{align}
One can use (\ref{eq:general_frame}) along with the equation of state, \eqref{eq:EOS}, to compute the temperature
\begin{equation} \label{eqn:temp_frame}
T = \frac{\Gamma - 1}{n} (T^{tt} - m n) - \tau_{\epsilon} \dot{T}
\end{equation}
which clearly illustrates that
different choices of hydrodynamic frame (i.e. values of $\tau_{\epsilon}$) lead to
different temperatures outside of equilibrium (in this case when $\dot{T} \neq 0$).
Thus the notion of temperature outside of equilibrium is intrinsically frame-dependent, 
whereas the total energy content, $T^{tt}$, is frame-independent;
as a result, one is able to excite dynamics which are ``purely frame'', even when
the system's physical observables ($T^{ab}, J^{a}$) are static.

One can also see from (\ref{eqn:temp_frame}) that removing the temperature's frame 
dependence, as is done in Eckart theory by taking $\tau_{\epsilon} \to 0$, results 
in a violation of the causality constraint (\ref{eq:simple_constraints}).
The BDNK and MIS
solutions in (\ref{eq:theory_comp}) provide the natural causal
generalization of the Eckart behavior, and relax $\epsilon \to T^{tt}$ with
timescale $\tau$.  In the special case where $\tau$ is independent of $t$,
one can integrate the BDNK/MIS results from (\ref{eq:theory_comp}) directly
to find
\begin{equation} \label{eq:exp_relax}
\epsilon(t) = T^{tt} + (\epsilon_{0} - T^{tt}) e^{-\frac{t}{\tau}},
\end{equation}
where $T^{tt}$ and $\epsilon_{0} \equiv \epsilon(t = 0)$ are freely
specifiable as initial data.  Note that initial data with $\epsilon_{0}
\neq T^{tt}, \dot{\epsilon} \neq 0$ has $\epsilon \to T^{tt}$ exponentially
fast, with timescale set by $\tau$. 
Specifying $\epsilon_0$
then corresponds to the choice of initial out-of-equilibrium contribution
$\delta \epsilon$. Since there is no physical dissipation
\textit{driving} the out-of-equilibrium corrections, the choice of
hydrodynamic frame leads to an exponential decay of this correction,
which recovers the perfect fluid value at late times. Causality is then
maintained by enforcing that this relaxation cannot occur too quickly.

It may also be somewhat surprising that BDNK theory gives a relaxation
equation in (\ref{eq:theory_comp}) like MIS theory, which is constructed to
obey relaxation-type PDEs.  
It is clear from (\ref{eq:theory_comp}) that
Eckart theory does not possess a relaxation-type structure, so it must come
from the additional terms added to Eckart theory to obtain BDNK theory.
These terms are proportional to
\begin{align}
u_{a} \nabla_{b} T^{ab}_{0} &= \big[ u^b \nabla_b \epsilon + \rho \nabla_b u^b \big] \label{eq:scalar_reg_term}  \\
\Delta^{a}_{~c} \nabla_{b} T^{bc}_{0} &= \big[ \rho u^{b} \nabla_{b} u^{a} + \Delta^{a b} \nabla_{b} P \big], \label{eq:vector_reg_term}
\end{align}
where (\ref{eq:scalar_reg_term}) is added to the energy
density and the pressure, and (\ref{eq:vector_reg_term}) is
added to the heat flux vector prior to the application of a thermodynamic
identity; see \cite{Bemfica:2020zjp} or Footnote
\ref{footnote:Eckart_heat_flux}.  Though each term only possesses first
gradients of the hydrodynamic variables, since it is proportional to
the relativistic Euler equations $\nabla_{a} T^{ab}_{0}$, it is equal to
zero up to first order in gradients and thus is of
$\mathcal{O}(\nabla^{2})$ \textit{on-shell}, i.e. when evaluated on
solutions to the equations of motion.  As a result, adding the terms
(\ref{eq:scalar_reg_term}-\ref{eq:vector_reg_term}) to the Eckart
stress-energy tensor does not alter it up to $\mathcal{O}(\nabla)$, and for
this reason BDNK theory is said to have the same ``physical content'' as
Eckart theory.  The terms
(\ref{eq:scalar_reg_term}-\ref{eq:vector_reg_term}) are responsible for the
relaxation form of (\ref{eq:theory_comp}).  This can be shown in general by
computing, for example, ${u_{a} u_{b} T^{ab}}$ from
(\ref{eq:Tab},\ref{eq:script_E}), which can be rearranged to yield
\begin{equation}
\begin{aligned}
u^c \nabla_c \epsilon =& \frac{1}{\tau_{\epsilon}} (u_{a} u_{b} T^{ab} -
\epsilon) - \rho \nabla_c u^c \,, \\
=& \frac{1}{\tau_{\epsilon}} \delta \epsilon - \rho \nabla_c u^c \,,\,
\end{aligned}
\end{equation}
where to obtain the second line we have used \eqref{eq:general_frame}. 
The first line is a relaxation equation (cf. (\ref{eq:MIS_relax})),
and the second line is equal to the energy conservation
equation for the perfect fluid, (\ref{eq:scalar_reg_term}),
with the addition of the out-of-equilibrium correction term $\delta
\epsilon$.

Purely frame dynamics and the relaxation form of the BDNK equations will be
essential to the discussions of the tests in the following sections, each
of which will consider scenarios with symmetries far more relaxed than in
the simple example considered in this section.

\subsection{(0+1)D uniform expansion: Bjorken flow} \label{sec:Bjorken_flow}

In this section we investigate the impact of the choice of the relaxation times $\tau_{\epsilon}, \tau_{P}, \tau_{Q}$ on solutions to the BDNK equations in a dynamical setting.  In particular, we aim to address the following questions: (1) does the behavior of the solution \textit{qualitatively} change when dissipation is applied too quickly (such that the characteristic speeds are superluminal)? And (2) what happens in the opposite limit, when dissipation is applied too slowly?

A natural starting point is to consider the simplest possible system with time dynamics, so we will specialize to boost-invariant uniformly expanding flows as were famously described by Bjorken \cite{Bjorken1983}.  Boost invariance is defined by the flow having a velocity profile which remains unchanged upon Lorentz boosts along a given axis. In Cartesian coordinates $x^{a} = (t, x, y, z)$, such a profile can be described by a three-velocity of the form $v^{z} = z/t$ (where we have defined the boost-invariant axis to be $z$). 

Bjorken flow is not typically studied in Cartesian coordinates, however, as the Cartesian equations of motion take the form of coupled PDEs.  If one instead uses Milne coordinates $x^{a} = (\tau, x, y, \xi)^{T}$, where the proper time coordinate is $\tau \equiv \sqrt{t^2 - z^2}$ and the rapidity is $\xi \equiv \textnormal{arctanh}(z/t)$, the equations of motion reduce to a single ODE; for this reason we study Bjorken flow in Milne coordinates.  Minkowski spacetime in Milne coordinates is characterized by the metric
\begin{equation}
g_{ab} = \textnormal{diag}(-1, 1, 1, \tau^2),
\end{equation}
which leads to nonzero Christoffel symbols
\begin{equation}
\Gamma^{\xi}_{\tau \xi} = \Gamma^{\xi}_{\xi \tau} = \frac{1}{\tau}, ~~~ \Gamma^{\tau}_{\xi \xi} = \tau,
\end{equation}
and the square root of the metric determinant is ${\sqrt{|g|} = \tau}$.  Bjorken flow assumes that there are no dynamics transverse to the $z$-axis, so everything is independent of $x, y, \xi$ and $u^{a} = (1, 0, 0, 0)^{T}$.  

The particle current conservation law (\ref{eq:Ja_cons_law}) then immediately implies
\begin{equation}
n(\tau) = \frac{n_0}{\tau},
\end{equation}
where $n_0$ is a spacetime constant.  Only one component of the stress-energy conservation law (\ref{eq:Tab_cons_law}) is nontrivial due to the high degree of symmetry, the $\tau$ component, which can be written
\begin{equation} \label{eq:Bjorken_EOM}
\tau_{\epsilon} \ddot{\epsilon} = -\frac{1}{\tau} (\tau + 2 \tau_{\epsilon} + \tau_{P}) \dot{\epsilon} - \frac{1}{\tau^2} \big[ \rho (\tau+\tau_{P}) - V \big]
\end{equation}
where $\ddot{\epsilon} \equiv \partial_{\tau}^{2} \epsilon$, $\dot{\epsilon} \equiv \partial_{\tau} \epsilon$, and the transport coefficients $\{\tau_{\epsilon}, \tau_{P}, V\}$ are defined in (\ref{eq:hydro_frame}-\ref{eq:V}).  As mentioned in Sec. \ref{sec:results}, the high degree of symmetry and the flat spacetime background imply the shear and bulk viscosities only appear in the combination $V$, and the heat flux vanishes, $\mathcal{Q}^{a} = 0$, so the thermal conductivity $\sigma$ and relaxation time $\tau_{Q}$ do not appear.

Before investigating the dynamics of the BDNK solutions, it is useful to first orient oneself using the corresponding inviscid solution, which for a fluid with the relativistic ideal gas equation of state (\ref{eq:EOS}) is given by the $\tau_{\epsilon}, \tau_{P}, V \to 0$ limit of (\ref{eq:Bjorken_EOM}) which has solution
\begin{equation} \label{eq:inviscid_bjorken}
\epsilon(\tau) = m n_0 \tau^{-1} \big[ 1 + e_0 \tau^{-(\Gamma - 1)} \big] ~~~ \textnormal{(inviscid)},
\end{equation}
where $e_0$ is an integration constant. Comparison of the inviscid solution
(\ref{eq:inviscid_bjorken}) with the definition of the specific internal
energy density $e$ (\ref{eq:e_defn}) shows that $\epsilon$ separates into a
rest mass energy density term $\sim m n_0$ which decays like $1/\tau$, and
an internal energy density term which decays more quickly since $1 < \Gamma
< 2$.  The local characteristic speeds of the relativistic Euler equations
(at zero fluid velocity) are always equal to the local sound speed
$c_s$ (\ref{eq:cs_sq}), which is bounded above by unity and thus solutions
are always causal for reasonable choices of the equation of state (such as
(\ref{eq:EOS}) with $1 < \Gamma < 2$). 

\begin{figure}
\centering
\includegraphics[width=\columnwidth]{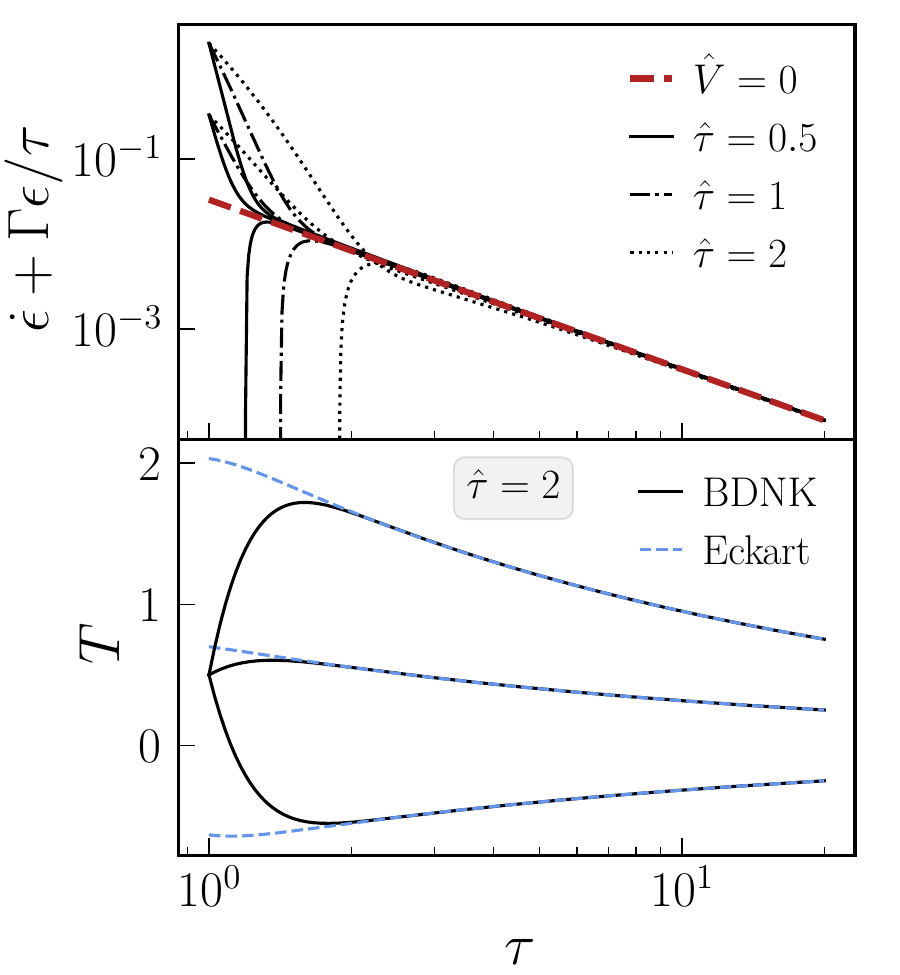}
\caption{Top panel: plot of the quantity $\dot{\epsilon} + \Gamma \epsilon/\tau$ (equal to $m n_0 (\Gamma - 1)/\tau^2$ for the inviscid solution (\ref{eq:inviscid_bjorken}), independent of $e_0$) for BDNK solutions with $n_0 = 0.1$, $\epsilon_0 = 0.25$ and $\dot{\epsilon}_0 \in \{-2, 0, 2\}$ (other parameters held constant are listed in Table \ref{table:parameters}).  In solid, dot-dash, and dotted lines the relaxation time $\hat{\tau} \, (\propto \tau_{\epsilon})$ is varied between $0.5$ and $2$, the only noticeable effect of which is that larger relaxation times result in solutions which equilibrate (reach the inviscid solution, red dashed line) more slowly.  The presence of superluminal characteristic speeds has no noticeable impact on the solution, as the $\hat{\tau} = 0.5$ case's characteristics are always superluminal, $\hat{\tau} = 1$ has superluminal characteristics only at early times, and $\hat{\tau} = 2$ has strictly subluminal characteristics, and all three have the same qualitative behavior.
Bottom panel: the $\hat{\tau} = 2$ solutions ($\dot{\epsilon}_0 \in \{-2, 0, 2\}$) from the top panel, except plotted is the temperature $T$ in the BDNK frame (black solid lines) and a temperature which is computed in the Eckart frame at each $\tau$ (blue dashed lines; see the text for how this is computed).  Note that the bottom-most pair of lines exhibit $T < 0$ as a result of taking far-from-equilibrium initial data.  At large $\tau$ (beyond what is shown in the plot), $T > 0$ in both frames.
} \label{fig:bjorken}
\end{figure}

Moving on to the viscous theory, one can show that the BDNK characteristic
speeds for the relativistic ideal gas (\ref{eq:cpmsq}-\ref{eq:c1sq}) all
diverge in the limit ${\hat{\tau} ~(\propto \tau_{\epsilon}) \to 0}$.
Inspection of (\ref{eq:Bjorken_EOM}) in the limit 
$\hat{\tau} \to 0$
(keeping everything else, 
i.e. $\tau_{P}, V$
finite)
does not show any clear indication that the qualitative behavior of the solution will change when the characteristic speeds become superluminal.  This notion is supported by the numerical results below---see Fig. \ref{fig:bjorken}. 

The opposite limit 
$\hat{\tau} \to \infty$ (with $\tau_{P}, V$ finite),
however, reduces (\ref{eq:Bjorken_EOM}) to $\ddot{\epsilon} = -2 \dot{\epsilon}/\tau$ which can be analytically integrated to yield
\begin{equation}\label{eq_tau_inf}
\lim\limits_{\tau_{\epsilon} \to \infty} \epsilon(\tau) = c_1 \tau^{-1} + c_2.
\end{equation}
Note that this solution can, in principle, agree with the inviscid solution (\ref{eq:inviscid_bjorken}) in the limit $\tau \to \infty$ if $c_1 = m n_0, c_2 = 0$, but it will not agree in the ultrarelativistic limit ($m \to 0$ while keeping $m n_0 e_0$ finite) where the inviscid solution becomes $\epsilon \propto \tau^{-(\Gamma - 1)}$.  This result agrees with the intuition that the system should not equilibrate in the limit of infinite relaxation times.

To check the reasoning derived from the 
$\hat{\tau} \to 0, \infty$
limits described above, we numerically integrate (\ref{eq:Bjorken_EOM})
starting from initial data $\epsilon = 0.25$, $\dot{\epsilon} \in \{-2, 0,
2\}$ from $\tau = 1$ to $\tau = 20$ for a fluid with $n_0 = 0.1$ (other
parameters held constant are listed in Table \ref{table:parameters}).
These results are shown in the top panel of Fig. \ref{fig:bjorken}, which shows 
$\tau$ versus the quantity
$\dot{\epsilon} + \Gamma \epsilon/\tau$, which
evaluates to $m n_0 (\Gamma - 1)/\tau^2$ when evaluated on the inviscid
solution (\ref{eq:inviscid_bjorken}), independent of $e_0$.  In the plot,
one can see that the choice of relaxation time $\tau_{\epsilon} \propto
\hat{\tau}$ only impacts how long it takes the viscous solution to approach
the inviscid solution (red dashed line).  In the figure, the case
$\hat{\tau} = 0.5$ always has superluminal characteristics, as maximum
characteristic speed $c_+ \approx 1.3$; the $\hat{\tau} = 1$ case has
superluminal characteristics at early times ($c_+ \approx 1.05$) but
they are all subluminal at late times ($c_+ \approx 0.9$); and the
case with $\hat{\tau} = 2$ has characteristics which are always
subluminal, $c_+ \approx 0.7$.  There appears to be no qualitative change
in the solution when the characteristics are superluminal, aside from
approaching the inviscid solution more rapidly.  Numerically, however,
decreasing $\hat{\tau} \propto \tau_{\epsilon}$ makes the equation of
motion (\ref{eq:Bjorken_EOM}) a ``stiff'' ODE, requiring very small steps
in $\tau$ to still resolve the decay time $\tau_\epsilon$.

This behavior is also in line with the loose analogy between gauge
(coordinate) dynamics in general relativity and frame dynamics here. In the
former, gauge modes are not confined to propagate at the speed of
light and can be superluminal without leading to any causality violation.
Presumably here something similar holds, in that as long as physical
degrees of freedom do not propagate superluminally, that the underlying
PDEs have superluminal characteristics is not {\em a priori} a problem. Of
course, this (0+1)D situation by construction does not allow propagating
features relative to the Milne spatial coordinates, and any conclusions
drawn from it relating to the influence of superluminal characteristics on
the solutions are limited. Though in the following section we further
explore this issue in a (1+1)D setting, and find similar conclusions, at
least up to ``moderately'' superluminal frames.

We now turn to the evolution of the temperature,
motivated by the fact that its value outside of equilibrium is dependent
on the choice of hydrodynamic frame (see the discussion around 
(\ref{eqn:temp_frame})).
In the bottom panel of Fig. \ref{fig:bjorken}, we show the temperature
evolution for the $\hat{\tau} = 2$ solutions from the top panel, in solid
black lines.  Apparent from the figure is that one of the solutions has a
negative temperature and, therefore, also negative equilibrium 
pressure, $P<0$. Although this might seem puzzling at first, 
the causality and linear stability constraints of
BDNK theory only require that $\rho > \eta/\tau_Q$ \cite{Bemfica:2020zjp},
which in our case translates to
\begin{equation}\label{eq:Pmin}
P > \frac{\Gamma-1}{\Gamma} \left(\frac{\eta}{\tau_Q} - m n\right)\,,
 \end{equation}
which can easily become negative for small shear viscosities $\eta$, large relaxation times $\tau_{Q}$, or large rest-mass energy densities $m n$.  The only ``purely frame'' quantity of these three is $\tau_{Q}$, so one may decide to require an additional constraint on the choice of hydrodynamic frame,
\begin{equation}\label{eqn:tau_phys}
\tau_Q > \frac{\eta}{m n} \,.
\end{equation}
We stress again that this additional constraint is not required for
causality, but merely to ensure that $P$, defined to be the \textit{equilibrium} pressure, is positive for
the ideal fluid equation of state considered here 
(and we do not enforce (\ref{eqn:tau_phys}) in this work).
This again highlights the
ambiguity in defining hydrodynamic frames, as the hydrodynamic variables (including
$T$) are not unique outside of equilibrium and are only constrained in that
they must agree with thermodynamic observables upon equilibration.  The
difficulty, however, is that one typically uses the \textit{equilibrium}
equation of state to close the system of hydrodynamics equations, which
assumes equilibrium values for parameters such as the temperature (i.e.
that $T > 0$).  Here this does not cause mathematical problems, as the
equation of state (\ref{eq:EOS}) has a simple analytic form valid for $T <
0$, but in many cases of interest it is tabulated and simply does not exist
for negative temperatures, energy densities, etc., which are in principle
allowed during the evolution.

One possible, partial solution to the aforementioned problem
would be to take advantage of the non-uniqueness of the hydrodynamic
parameters outside of equilibrium to map between the hydrodynamic frame
used in the evolution scheme and the one where the equilibrium equation
of state is known. This would at least allow a more self-consistent
implementation of the desired microphysical equation of state (though of
course will not address how the latter may change outside of equilibrium).
We illustrate this in Fig. \ref{fig:bjorken} by computing $T^{\tau \tau}$,
which is a frame-independent physical observable, and then using its value
to compute an Eckart frame temperature (dashed blue lines).  This is done
using the definition $T^{\tau \tau}_{E} = \epsilon_{E}$, where
$\epsilon_{E}$ is the Eckart frame energy density, which can then be used
to compute the Eckart temperature $T_{E}$ using the equation of state
(\ref{eq:EOS}) after noting that the baryon density $n$ is identical
between the two frames.  One still finds that the bottom-most solution has
$T_{E} < 0$, owing to the fact that the initial data in that case is very
far from equilibrium (far enough, in fact, that it violates the weak energy
condition as $u_{a} u_{b} T^{ab} < 0$). Even though the top-most
solution has positive temperatures throughout, at early times it is also
far from equilibrium as judged by the difference between the BDNK-frame
and Eckart temperatures. This could be a further useful diagnostic (in
addition to the weak energy condition) to monitor when a solution might
be outside the realm of validity of first-order hydrodynamics.

We leave a detailed analysis of Bjorken flow for BDNK theory to a future work, and note that related analyses involving Bjorken flow, either in the context of kinetic theory in the general frame formalism or with conformal BDNK theory, are given in \cite{Bemfica_2018,Das:2020gtq,Das:2020fnr,Shokri:2020cxa,Das:2020grz,Noronha:2021syv,Rocha:2021lze,Rocha:2022ind}.

\subsection{(1+1)D approach to steady-state shockwave solutions} \label{sec:shockwaves}

\begin{figure}
\centering
\includegraphics[width=\columnwidth]{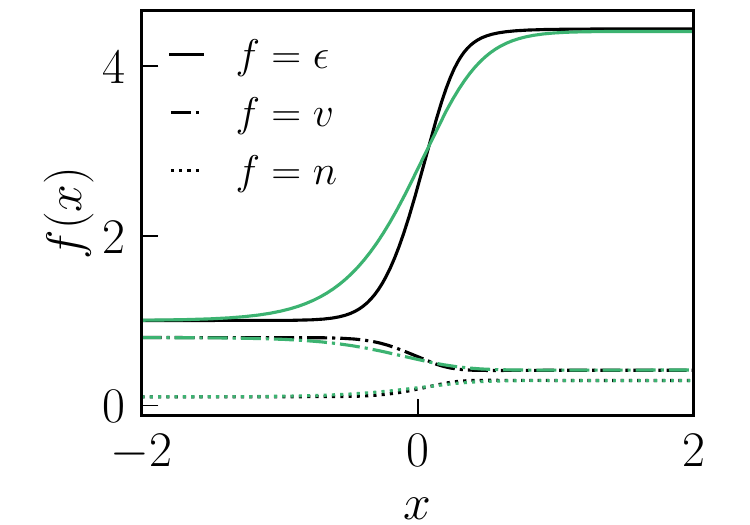}
\caption{Steady-state shockwave solutions to (\ref{eq:shockwave_nprime},
\ref{eq:shockwave_epsP}-\ref{eq:shockwave_velP}) for a BDNK fluid with
relativistic ideal gas microphysics (\ref{eq:EOS}) and the chosen frame
(\ref{eq:hydro_frame}) 
in black, and a conformal fluid with the same shear viscosity and
the sharply causal hydrodynamic frame (equation 16, frame {\tt B} of \cite{Pandya_2021}) in green.  For each case we use
asymptotic left state $\{\epsilon, v, n\} =
\{1, 0.8, 0.1\}$ as ${x \to -\infty}$; the other required parameters are
given in Table \ref{table:parameters}. 
The inviscid solution in each case would have the same asymptotic left and right states, except with the smooth transition between those states replaced with a step function discontinuity at $x \approx 0$.
} \label{fig:shockwave_profile}
\end{figure}

\begin{figure}
\centering
\includegraphics[width=\columnwidth]{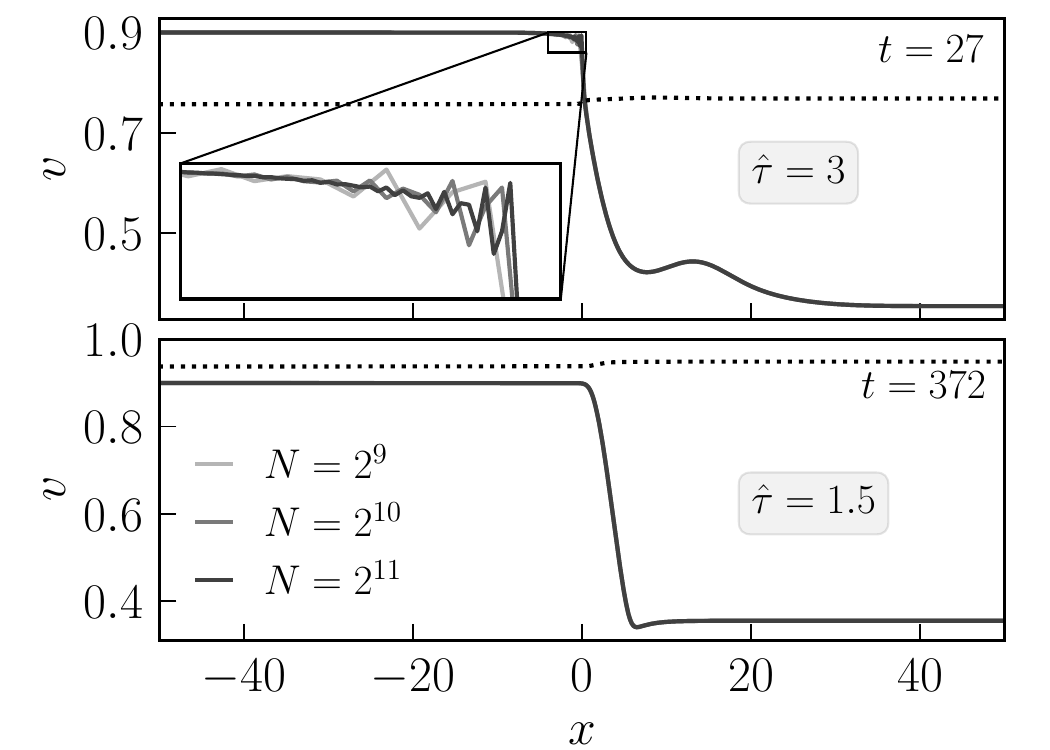}
\caption{Solutions starting from initial data (\ref{eq:shockwave_ID}, \ref{eq:shockwave_params}) for a BDNK fluid with ideal gas microphysics (\ref{eq:hydro_frame}) and parameters given in Table \ref{table:parameters}.  The top panel illustrates the onset of a high-frequency numerical instability in the region of the flow where the flow velocity $v$ (solid lines with darkness indicating greater numerical resolution) exceeds the maximum characteristic speed $c_+$ (dotted line).  In the bottom panel, a different hydrodynamic frame is chosen such that $c_+ > v$ throughout the shockwave, and no instability sets in (note the much later timestamp in the top-right corner).} \label{fig:shock_instability}
\end{figure}

\begin{figure}
\centering
\includegraphics[width=\columnwidth]{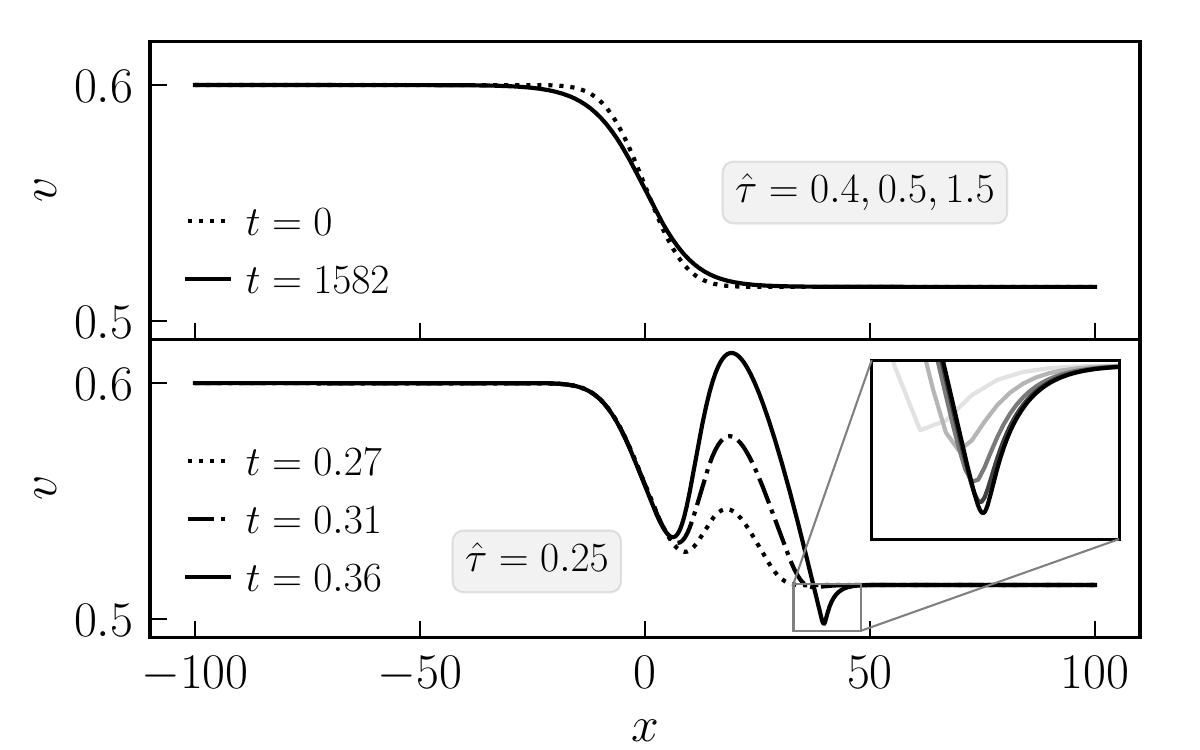}
\caption{Comparison of solutions for initial data (\ref{eq:shockwave_ID}, \ref{eq:shockwave_params}) for various hydrodynamic frames.  In the top panel, the solution is shown at $t = 0$ (dotted line) and at a very late time ($t = 1582$) when the dynamics have damped out and the shockwave very closely approximates the steady-state solution (solid line).  At each of these times, the solution for the three hydrodynamic frames with $\hat{\tau} = 0.4, 0.5, 1.5$ (corresponding to maximum characteristic speeds $c_+ \sim 1.6, 1.5, 0.9$) are identical up to the resolution of the plot.  The fact that the $\hat{\tau} = 0.5$ case has superluminal characteristics has no noticeable impact on the solution, though dropping much below $\hat{\tau} = 0.5$ results in the equations becoming ``stiff'' in the same manner as occurs for Bjorken flow, and very small timesteps are required to integrate them (the $\hat{\tau} = 0.4$ case requires CFL number $\lambda = 0.01$ for stable evolution, an order of magnitude smaller than that used in the other cases).  For wildly superluminal frames---one example of which is shown in the bottom panel, with $\hat{\tau} = 0.25 \implies c_+ \sim 2$---a very fast instability sets in at early times, stemming from the development of a large bump near $x \sim 20$ and a sharp feature ($x \sim 40$) shown in the inset (with numerical resolutions $N \in \{2^{7}, 2^{8}, 2^{9}, 2^{10}, 2^{11}\}$ in increasingly dark shades of gray).} \label{fig:acaus_instab}
\end{figure}

In this section we consider planar shockwave solutions, generalizing the analysis done in \cite{Pandya_2021} for the conformal BDNK fluid.  All of the results below, with the exception of the numerical solutions depicted in Figs. \ref{fig:shockwave_profile}-\ref{fig:acaus_instab}, are obtained without specifying a hydrodynamic frame or even the equation of state, and thus apply to BDNK fluids in general. 

Starting with a traveling planar shockwave solution in 4D Minkowski spacetime, one can boost into the rest frame of the shock so that the solution becomes independent of the time coordinate $t$; furthermore, one can align the (Cartesian) coordinate system such that all variation occurs in the $x$ direction. We choose to write the four-velocity $u^{a}$ in terms of the three-velocity in the $x$-direction, $v \in [0,1)$, as well as the Lorentz factor ${W \equiv (1 - v^2)^{-1/2}}$ via
\begin{equation}
u^{a} = (W, W v, 0, 0)^{T},
\end{equation}
which is timelike and thus satisfies $u_{c} u^{c} = -1$.  

The equations of motion (\ref{eq:Tab_cons_law}-\ref{eq:Ja_cons_law}) then reduce to ODEs for the three quantities $n(x), \epsilon(x), v(x)$.  Of these, the particle current conservation law (\ref{eq:Ja_cons_law})---which reduces to $(J^{x})' = 0$, where a prime represents $\partial_{x}$---is the simplest and yields
\begin{equation} \label{eq:shockwave_nprime}
n' = - \frac{W^2 n v'}{v},
\end{equation}
allowing one to eliminate $n'$ in favor of $v'$ in the other equations of motion.  The remaining two equations---$(T^{tx})' = 0$ and $(T^{xx})' = 0$---derive from (\ref{eq:Tab_cons_law}) and can be rearranged to isolate $\epsilon'(x), v'(x)$.  Doing so yields equations of the form
\begin{equation} \label{eq:shared_den}
\epsilon'(x), v'(x) \propto \frac{1}{A v^4 + v^2 ( B - \tau_{\epsilon} \delta )
+ ( C + \tau_{P} \delta )},
\end{equation}
where $A, B, C$ are defined in (\ref{eq:A}-\ref{eq:C}) and
\begin{equation}
\delta \equiv \beta_{\epsilon} \rho + \beta_{n} n - \rho c_s^2 \tau_{Q} - \sigma \kappa_s = 0
\end{equation}
which vanishes identically after inserting the definitions of
$\beta_{\epsilon}, \beta_{n}, c_s^2$, and $\kappa_{s}$,
(\ref{eq:beta_eps}-\ref{eq:kappa_s}).  Thus the shared denominator in
(\ref{eq:shared_den}) is quadratic in $v^2$ with roots
\begin{equation} \label{cpmsq_general}
c_{\pm}^2 = \frac{-B \pm \sqrt{B^2 - 4 A C}}{2 A},
\end{equation}
where $c_{\pm}$ are two of the three unique characteristic speeds of the BDNK equations.  In terms of $c_{\pm}$, the full system specifying a steady-state BDNK planar shockwave is given by the baryon current conservation equation (\ref{eq:shockwave_nprime}) in addition to
\begin{align}
\epsilon'(x) = \frac{c_4 v^4 + c_3 v^3 + c_2 v^2 + c_1 v + c_0}{A W v (v - c_{+})(v + c_{+})(v - c_{-})(v + c_{-})} \label{eq:shockwave_epsP} \\
v'(x) = \frac{d_3 v^3 + d_2 v^2 + d_1 v + d_0}{A W^3 (v - c_{+})(v + c_{+})(v - c_{-})(v + c_{-})} \label{eq:shockwave_velP} 
\end{align}
where the coefficients $c_i, d_i$ are given by
\begin{equation}
\begin{aligned}
&c_0 = \beta_{n} n (T^{xx} - P), ~~ c_1 = -T^{tx} (2 \beta_{n} n-\rho  \tau_{P}+V) \\
&c_2 = [\beta_{n} n - \rho (\tau_{\epsilon}+\tau_{P}+\tau_{Q}) + V] (T^{xx}+\epsilon) + \rho^2 (\tau_{\epsilon}+\tau_{Q}), \\
&c_3 = \rho  T^{tx} (\tau_{\epsilon}+2 \tau_{Q}), ~~~ c_4 = -\rho \tau_{Q} (T^{xx}+\epsilon) \\
&d_0 = \beta_{\epsilon} (T^{xx} - P), ~~~ d_1 = -T^{tx} (2 \beta_{\epsilon}+\tau_{P})\\
&d_2 = (T^{xx} + \epsilon) (\beta_{\epsilon}+\tau_{\epsilon}+\tau_{P}) - \rho  \tau_{\epsilon}, ~~~ d_3 = - \tau_{\epsilon} T^{tx}, \\
\end{aligned}
\end{equation}
and the quantities $T^{tx}, T^{xx}$ are constant by the equations of motion.

From inspection of (\ref{eq:shockwave_nprime}, \ref{eq:shockwave_epsP}-\ref{eq:shockwave_velP}) one immediately notices that the denominators of the three equations vanish for $v = \pm c_{\pm}, 0$, which will lead the system to have no solution unless the numerators simultaneously vanish.  It is difficult to verify whether the numerators vanish in general if $v$ attains these values, since the numerator of (\ref{eq:shockwave_epsP}) is a fourth-order polynomial in $v$ and (\ref{eq:shockwave_velP}) is one of third order, implying that the general form of the roots will be quite complicated.  Instead of evaluating the roots analytically, we will implicitly try to understand this structure by 
numerically solving the BDNK fluid equations with ideal gas microphysics.

The solution to the system of coupled ODEs (\ref{eq:shockwave_nprime},
\ref{eq:shockwave_epsP}-\ref{eq:shockwave_velP}) is shown in Fig.
\ref{fig:shockwave_profile} for a pair of shockwaves each with asymptotic left state
$\{\epsilon_{L}, v_{L}, n_{L}\} = \{1, 0.8, 0.1\}$ as ${x \to -\infty}$.
In black, we show the shockwave profile for a BDNK fluid with ideal gas microphysics
(with other parameters in Table \ref{table:parameters}) and in green a conformal
BDNK fluid as studied in \cite{Pandya_2021}.
The
two required (constant) stress-energy tensor components $T^{tx}, T^{xx}$
appearing in (\ref{eq:shockwave_epsP}-\ref{eq:shockwave_velP}) are computed
using the perfect fluid stress-energy tensor since the asymptotic states at
${x \to \pm \infty}$ should be in thermodynamic equilibrium, so
(\ref{eq:Tab_0}) implies $T^{tx} = \rho W^2 v$ and $T^{xx} = \rho W^2 v^2 +
P$ and these values are computed using $\epsilon_{L}, v_{L}, n_{L}$.  
For both shockwaves shown in Fig. \ref{fig:shockwave_profile}, the velocity
profile never
attains a value which makes the denominators in
(\ref{eq:shockwave_epsP}-\ref{eq:shockwave_velP}) vanish.  Using an
asymptotic left state with $v_L \geq c_+$ results in the solver finding the
trivial equilibrium state where $\epsilon, v, n = \epsilon_{L}, v_{L},
n_{L}$ are all constant $\forall x$. 

To address what occurs when a shockwave forms dynamically, we follow \cite{Pandya_2021} in using a set of initial data which approximates the profile of a shockwave in its rest frame:
\begin{equation} \label{eq:shockwave_ID}
\begin{aligned}
\epsilon(0,x) &= \frac{\epsilon_{R} - \epsilon_{L}}{2} \Big[ \textnormal{erf}\Big(\frac{x}{w}\Big) + 1 \Big] + \epsilon_{L} \\
v(0, x) &= \frac{v_{L} - v_{R}}{2} \Big[ 1 - \textnormal{erf} \Big( \frac{x}{w} \Big) \Big] + v_{R} \\
n(0, x) &= \frac{n_{L} - n_{R}}{2} \Big[ 1 - \textnormal{erf} \Big( \frac{x}{w} \Big) \Big] + n_{R}, \\
\end{aligned}
\end{equation}
where $\textnormal{erf}(y)$ is the Gaussian error function and each of the above functions is chosen to smoothly interpolate between the asymptotic left states obtained as $x \to -\infty$ ($\epsilon_{L}, v_{L}, n_{L}$) and the asymptotic right states obtained as $x \to \infty$ ($\epsilon_{R}, v_{R}, n_{R}$); the width of the transition is controlled by $w$.  We treat the left states as freely specifiable, and fix the right states using the Rankine-Hugoniot conditions for a shockwave in its rest frame,
\begin{equation} \label{eq:Rankine_Hugoniot}
\begin{aligned}
n_{L} W_{L} v_{L} &= n_{R} W_{R} v_{R} \\
v_{L} W_{L}^2 \rho_{L} &= v_{R} W_{R}^2 \rho_{R} \\
v_{L}^2 W_{L}^2 \rho_{L} + P_{L} &= v_{R}^2 W_{R}^2 \rho_{R} + P_{R},
\end{aligned}
\end{equation}
where $W_{i} = (1 - v_{i})^{-1/2}$, $\rho_{i} = \epsilon_{i} + P_{i}$, and we solve the above relations numerically to find the left-right state pairs $\{\epsilon, v, n\}$
\begin{equation} \label{eq:shockwave_params}
\begin{aligned}
\{ 1, 0.9, 1 \}_{L} &\implies \{ 11.5174, 0.354727, 5.44212 \}_{R} \\
\{ 1, 0.6, 1 \}_{L} &\implies \{ 1.33795, 0.514414, 1.25027 \}_{R} \\
\end{aligned}
\end{equation}
which are used in Fig. \ref{fig:shock_instability} and Fig. \ref{fig:acaus_instab} respectively.

A comparison of states from these initial data (\ref{eq:shockwave_ID},
\ref{eq:shockwave_params}) with $w = 10$ and different hydrodynamic frames
(${\hat{\tau} = 1.5, 3}$) are shown in Fig. \ref{fig:shock_instability} for a
BDNK fluid with ideal gas microphysics (other parameters are listed in
Table \ref{table:parameters}).  As occurs in the case of a conformal fluid
(described in detail in \cite{Pandya_2021}), the evolution exhibits a
high-frequency numerical instability 
when part of the velocity profile
exceeds the maximum characteristic speed $c_+$, precisely the same case
where the ODEs describing a steady-state shockwave,
(\ref{eq:shockwave_nprime},\ref{eq:shockwave_epsP}-\ref{eq:shockwave_velP}),
yield no solution.  When a hydrodynamic frame is chosen such that $|c_+| >
|v|$ across the entire shockwave profile, the evolution is stable and at
late times asymptotes to a solution of the steady-state ODEs
(\ref{eq:shockwave_nprime},
\ref{eq:shockwave_epsP}-\ref{eq:shockwave_velP}). 
This intuition---namely that the corresponding steady-state solution
must exist for dynamical shockwave solutions to be stable---is consistent
with a mathematically rigorous result for conformal BDNK fluids
\cite{Freistuhler2021}, which showed that a given hydrodynamic frame will
always possess shockwave solutions which break down unless the maximum
local characteristic speed is greater than or equal to the speed of light.

One could similarly ask whether issues arise when the velocity profile
of the shockwave crosses $c_-$, the other nonzero root of the
denominators in (\ref{eq:shockwave_epsP}-\ref{eq:shockwave_velP}).  For
the hydrodynamic frame considered here, (\ref{eq:hydro_frame}), we find
empirically that one must severely violate both the causality and linear
stability constraints for such a case to occur.  We observe the onset of an
instability in those cases, but it is unclear whether it arises due to
nonexistence of the steady-state shockwave solution or due to an entirely
different mechanism triggered by violation of the aforementioned
constraints.  We leave further consideration of this case to a future
work.

In Fig. \ref{fig:acaus_instab} we examine the qualitative behavior of
shockwave solutions for the initial data
(\ref{eq:shockwave_ID},\ref{eq:shockwave_params}) with $w = 10$,
$\epsilon_{L}, v_{L}, n_{L} = \{1, 0.6, 1\}$ for hydrodynamic frames
with superluminal characteristics.  The figure shows that ``weakly''
superluminal frames, in this case with $c_+ \sim 1.5$
($\hat{\tau} = 0.5$)
do not present any issues, and give solutions which are identical
up to the resolution of Fig. \ref{fig:acaus_instab} to those produced
from the subluminal frame ${\hat{\tau} = 1.5 \implies c_+ \sim 0.9}$ (but
not exactly identical, as they converge to slightly different solutions in
the continuum limit).
Moreover, here, compared to the results discussed in the previous section,
we do have propagating dynamics. In particular notice the ``bump'' to the
right of the initial transition in the top panel of Fig.
\ref{fig:shock_instability}, which is sourced by the part of our initial
data (\ref{eq:shockwave_ID}) that deviates from the stationary shockwave
state, hence also forms in numerically stable cases, and propagates
downstream to the right at essentially the sound speed of the fluid. For
the frames with superluminal characteristics we do not see any
features propagating superluminally, nor even upstream. This gives further
evidence that superluminal characteristics are not {\em a priori} related
to physical propagation speeds, and hence do not necessarily lead to
causality violation.  That said, requiring all characteristics to be
subluminal \textit{guarantees} causality is respected, and thus such a
restriction should be considered an essential component of constructing a
sensible relativistic fluid theory.
One may even go so far as to require $|c_+| = 1$ (or, at the very least, $|c_+| = 1 - \delta$, for infinitesimal $\delta > 0$) to ensure that the system's characteristics are causal \textit{and} that fast shockwave solutions do not exhibit the instability displayed in Fig. \ref{fig:shock_instability}.

As the relaxation times are decreased, the equations become ``stiff''
numerically, meaning one must significantly reduce the CFL number for a
stable evolution (e.g. the case with $\hat{\tau} = 0.4 \implies c_+ \sim
1.6$ requires an order of magnitude smaller CFL number than $\hat{\tau} =
0.5 \implies c_+ \sim 1.5$) though one still recovers a solution at late
times which is effectively identical to the case with strictly
subluminal characteristics.  

For ``wildly'' superluminal frames, however, we observe the onset of a very fast instability, shown in the bottom panel of Fig. \ref{fig:acaus_instab}.
In this case, we find that as the initial transient ``bump'' begins to leave the shockwave, rather than growing to a fixed size and propagating away, it grows unboundedly without propagating.  The growth of this feature can be seen in the dotted, dot-dash, and solid lines in the bottom panel of Fig. \ref{fig:acaus_instab}.  Shortly later, the quantities $\dot{\epsilon}, \dot{v}$ appear to diverge in finite time at a point to the right of the bump.  This divergence forms a very sharp feature in essentially all state variables, though none of them---including $c_s, c_\pm, T^{ta}, \epsilon, P, n$---appear to obtain unphysical values as a result.  The sharp feature in the $v$ profile is shown in the inset at multiple numerical resolutions whose behavior appear to be indicating convergence.  This implies that the very rapid growth up until this point is present in the continuum PDE solution and is not numerical in origin.  Shortly beyond the time shown in Fig. \ref{fig:acaus_instab}, the sharp feature sources an oscillatory numerical instability which crashes the simulation.

To briefly summarize, we find that there is no qualitative change in the solution for ``weakly superluminal'' frames, but the equations become stiff numerically as the characteristic speeds increase.  At some point, when the fastest characteristic speed is much greater than unity an instability may set in, potentially leading to singular behavior in the solution. 
This instability is likely related to the fact that subluminal characteristics are required in the BDNK formalism's proof of linear stability.
We further investigate the stability of BDNK solutions in the next section, in the context of heat flow.

\subsection{(1+1)D heat conduction} \label{sec:heat_flow}

\subsubsection{Heat flow problem with constant coefficients}

In this section we consider heat flow in BDNK theory for a fluid with relativistic ideal gas microphysics, inspired in part by the analysis of \cite{Most:2021rhr} in the context of dissipative magnetohydrodynamics.  In particular, we will specialize to problems in (3+1)D Minkowski spacetime which vary only in time $t$ and one Cartesian spatial coordinate, $x$, and furthermore we restrict to solutions which never source a flow velocity, so $u^{i} = 0$ at all times (where $i$ is a spatial index).  Before working out the equations of motion, it is useful to apply the thermodynamic identity
\begin{equation} \label{eq:thermo_identity}
\frac{d P}{\rho} = \frac{d T}{T} + \frac{n T}{\rho} d(\mu/T)
\end{equation}
to the BDNK particle current to write it in the form
\begin{equation} \label{eq:alt_heat_vector}
\mathcal{Q}^{a} = - \kappa \Delta^{a b} \nabla_b T + \tau_{Q} \rho u^{b} \nabla_{b} u^{a} + \gamma \Delta^{a b} \nabla_{b} P,
\end{equation}
where  $\kappa \equiv \frac{\sigma \rho^2}{n^2 T}$ (\ref{eq:kappa}) and 
\begin{equation} \label{eq:gamma}
\gamma \equiv \tau_{Q} + \frac{\sigma \rho}{n^2}.
\end{equation}

Moving on to the equations of motion, the baryon current equation (\ref{eq:Ja_cons_law}) implies
\begin{equation} \label{eq:heat_baryon_EOM}
\dot{n} = 0,
\end{equation}
and the two nontrivial components ($t, x$) of (\ref{eq:Tab_cons_law}) can be written
\begin{align}
0 &= (\epsilon + \tau_{\epsilon} \dot{\epsilon})_{,t} + (- \kappa T' + \gamma P')_{,x} \label{eq:heat_t_eqn} \\
0 &= (- \kappa T' + \gamma P')_{,t} + (P + \tau_{P} \dot{\epsilon})_{,x}. \label{eq:heat_x_eqn}
\end{align}
Note that as written, the variables $T, P$ are each functions of the hydrodynamic variables $\epsilon, n$ and are related by the equation of state (\ref{eq:EOS}).

To investigate the causality and stability of ``pure heat flow'' solutions to the BDNK equations, we will now consider three different classes of hydrodynamic frames:
\begin{equation} \label{eq:heat_frames}
\begin{aligned}
&\textnormal{Eckart frame:}  &&\tau_{\epsilon}, \tau_{P} = 0, \tau_{Q} = -\frac{\kappa T}{\rho} \\
&\textnormal{Eckart/BDNK hybrid:} &&\tau_{\epsilon} > 0, \tau_{P} = 0, \tau_{Q} = -\frac{\kappa T}{\rho} \\
&\textnormal{BDNK frame:} &&\tau_{\epsilon}, \tau_{P}, \tau_{Q} > 0,
\end{aligned}
\end{equation}
where the choice of $\tau_{Q}$ in the first two frames implies $\gamma = 0$ and the pressure gradient vanishes from the heat flux vector (\ref{eq:alt_heat_vector}).  For the remainder of this subsection, to simplify calculations we will take the transport coefficients $\{\tau_{\epsilon}, \kappa, \gamma\}$ to be spacetime constants.  Note that though one is free to choose constant transport coefficients within the scope of Eckart and BDNK theories, this assumption is not true for the hydrodynamic frame chosen in Sec. \ref{sec:model}, which we will return to in the next subsection.

For each of the three hydrodynamic frames described in (\ref{eq:heat_frames}), the respective $t$-components of the equations of motion (\ref{eq:heat_t_eqn}) can be written entirely in terms of the variables $T, n$ in the form:
\begin{align}
0 &= \dot{T} - \alpha_E T'' &&\textnormal{(Eckart)}  \label{eq:heat_t_Eckart} \\
0 &= \ddot{T} - c_h^2 T'' + \frac{1}{\tau_{\epsilon}} \dot{T} &&\textnormal{(hybrid)} \label{eq:heat_t_hybrid} \\
0 &= \ddot{T} - c_B^2 T'' + \frac{1}{\tau_{\epsilon}} \dot{T} + l.o.t. &&\textnormal{(BDNK)} \label{eq:heat_t_BDNK}
\end{align}
where ${\alpha_E \equiv \frac{\kappa (\Gamma - 1)}{n}}$, ${c_h^2 \equiv \frac{\kappa (\Gamma - 1)}{n \tau_{\epsilon}}}$, ${c_B^2 \equiv c_h^2 (1 - \frac{\gamma n}{\kappa})}$, and the 
lower-order terms $l.o.t. \equiv \frac{(\Gamma - 1)}{n \tau_{\epsilon}} \gamma (n'' T + 2 n' T')$.
Inspection of (\ref{eq:heat_t_Eckart}-\ref{eq:heat_t_BDNK}) provides a clear physical illustration of how the choice of hydrodynamic frame impacts the causal properties of the equations of motion.  Beginning with the Eckart frame equation, (\ref{eq:heat_t_Eckart}), one can identify it as the heat equation with thermal diffusivity $\alpha_E$ (the definition of which depends on the equation of state).  The heat equation is a second-order parabolic PDE, and thus may be thought of as describing a system which propagates information infinitely fast, violating causality.  Causality may be restored, however, for frames like the hybrid Eckart/BDNK frame.  One can see from the hybrid frame's equation of motion, (\ref{eq:heat_t_hybrid}) that it is a second-order hyperbolic PDE with a wave-like principle part with ``thermal propagation speed'' $c_h^2$.  In particular, (\ref{eq:heat_t_hybrid}) is an example of the so-called telegrapher's equation, which is the natural hyperbolic generalization of the heat equation \cite{Chester1963}.  The BDNK frame's equation of motion, (\ref{eq:heat_t_BDNK}), is similar to that of the hybrid frame except it possesses a modified thermal propagation speed $c_B^2$ as well as additional lower-order terms.

As was discussed in Sec. \ref{sec:introduction}, Eckart theory is both acausal and linearly unstable about thermodynamic equilibrium, while BDNK theory is causal and stable.  We can address the cause of the instability for the heat flow problem by considering how the choice of hydrodynamic frame impacts the $x$-component of the stress-energy conservation law, (\ref{eq:heat_x_eqn}), which for constant $\{\tau_{\epsilon}, \kappa, \gamma\}$ becomes
\begin{equation} \label{eq:heat_x_eqn_ODE}
0 = (\theta \dot{T} + P)'
\end{equation}
where
\begin{equation} \label{eq:heat_theta_defn}
\theta = 
\begin{cases}
-\kappa & \textnormal{(Eckart, hybrid)} \\
-\kappa + \gamma n + \tau_{P} \frac{n}{(\Gamma - 1)} & \textnormal{(BDNK)}
\end{cases}
\end{equation}
and in each case $\theta$ is a constant in time (but not in space in the BDNK case, as $n = n(x)$ by (\ref{eq:heat_baryon_EOM})).  For all three classes of hydrodynamic frames (\ref{eq:heat_x_eqn_ODE}) may be integrated once trivially in $x$ yielding an integration constant-in-$x$ $P_0(t)$, then solve for $\dot{T}$, multiply each side by $n(x)$, and rearrange to yield
\begin{equation} \label{eq:heat_x_soln}
\dot{P} = \frac{1}{\tau_{\theta}} (P_0(t) - P). 
\end{equation}
This is a relaxation equation for the pressure with relaxation time $\tau_{\theta} \equiv \theta/n$.  Inspection of (\ref{eq:heat_x_soln}) provides a simple explanation for why the Eckart frame exhibits instability for the heat flow problem: the relaxation time $\tau_{\theta} = -\kappa/n$ is negative, forcing $P$ to exponentially deviate from $P_0$ rather than relax to it.  Since the hybrid Eckart/BDNK frame shares the same (negative) value for $\theta$, it is unstable in the same fashion.  The BDNK frame, on the other hand, can be stable as long as $\theta \geq 0$, which can be thought of as an upper bound on $\sigma$ (note that the BDNK linear stability constraints give a similar bound, (\ref{eq:simple_constraints}), for the frame (\ref{eq:hydro_frame}) with non-constant coefficients; see Sec. \ref{sec:model}).
We remark that the system (\ref{eq:heat_t_Eckart}-\ref{eq:heat_t_BDNK}), (\ref{eq:heat_x_soln}) may be too restrictive to have global solutions---in other words, the $u^{i} = 0$ constraint may need to be slightly violated in order for solutions to exist.  The analysis above should hold approximately in these cases, however, and we consider cases with $u^{i} \neq 0$ in the next subsection.

\subsubsection{Heat flow problem with non-constant coefficients}

In this subsection we consider a slightly relaxed heat flow problem: rather than asserting that $u^{i}$ always remain zero, we instead simply require that it be zero at $t = 0$.  To specialize to heat flow solutions, we adopt initial data which possesses a temperature gradient but no pressure gradients\footnote{It is impossible to pose data which have temperature gradients but no pressure gradients for a conformal fluid, as there $P \propto T^4$.  In this sense, heat flow for a conformal fluid is ``trivial'' in that it is always accompanied by flow due to the pressure gradient. \label{footnote:heat_flow}}, 
namely
\begin{equation} \label{eq:heat_flow_ID}
T(0, x) = A e^{-\frac{x^2}{w^2}} + \delta, ~~~ P(0, x) = P_0 = \textnormal{const}.
\end{equation}
Since the BDNK PDEs are formulated in terms of hydrodynamic variables $\epsilon, n$ rather than $T, P$, we implement the initial data (\ref{eq:heat_flow_ID}) using the relations ${\epsilon = P [m T^{-1} + (\Gamma - 1)^{-1}]}$, ${n = P T^{-1}}$, which are derived from the equation of state (\ref{eq:EOS}).  Furthermore, we choose time-symmetric initial data to specify the required first time derivatives, so $\dot{\epsilon}(0,x) = \dot{u}^{i}(0,x) = 0$.

\begin{figure}
\centering
\includegraphics[width=\columnwidth]{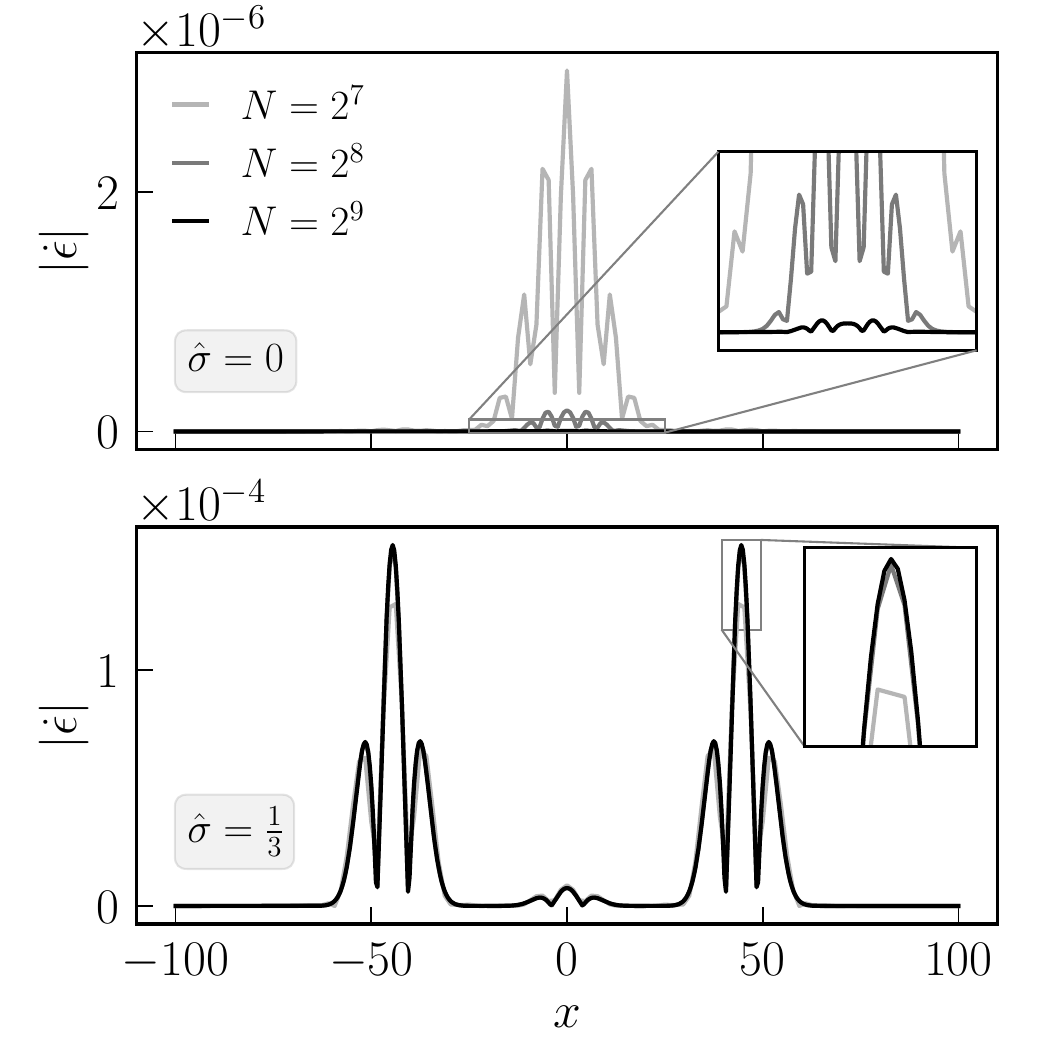}
\caption{Snapshots from an evolution starting from initial data (\ref{eq:heat_flow_ID}) for a BDNK fluid (see Table \ref{table:parameters} for a list of other parameters held constant) with $\hat{\sigma} = 0$ (top panel) as well as $\hat{\sigma} = 1/3$ (bottom panel) at a time shortly after $t = 0$ (darker gray for higher numerical resolution).  In the top panel, the fact that $\sigma = 0$ implies that $\ddot{\epsilon} = 0$ in (\ref{eq:heat_ID_EOM}), so the solution should not have dynamics; the fact that it does here is purely a result of numerical error, and $\dot{\epsilon}$ converges to zero as the grid is refined.  In the bottom panel, since $\sigma > 0$, there exists a dynamical heat flow solution and $\dot{\epsilon}$ converges to a nonzero value over part of the domain.} \label{fig:heat_stationary}
\end{figure}

For this set of initial data, (\ref{eq:Ja_cons_law}) again reduces to $\dot{n} = 0$, and one can straightforwardly show that the $x$-component of (\ref{eq:Tab_cons_law}) is trivially satisfied at $t = 0$, leaving just the $t$-component:
\begin{equation} \label{eq:heat_ID_EOM}
T^{at}_{,a} \Big|_{t=0} = 0 = \tau_{\epsilon} \ddot{\epsilon} - (\kappa T')'.
\end{equation}
Notice that the solution only has dynamics so long as $\kappa \neq 0$,
which in turn implies $\sigma \neq 0$ by (\ref{eq:kappa}); in other words,
the system only has dynamical heat flow solutions if the thermal
conductivity is nonzero.  We confirm that this statement is true in Fig.
\ref{fig:heat_stationary} for a BDNK fluid with $\hat{\sigma} = 0$ (top
panel) and $\hat{\sigma} = 1/3$ (bottom panel) and other constant
parameters listed in Table \ref{table:parameters}.  Notice that
$\dot{\epsilon}$ converges to zero with numerical resolution when $\sigma =
0$, and converges to a finite value for $\sigma \neq 0$. 

\begin{figure*}
\centering
\includegraphics[width=\textwidth]{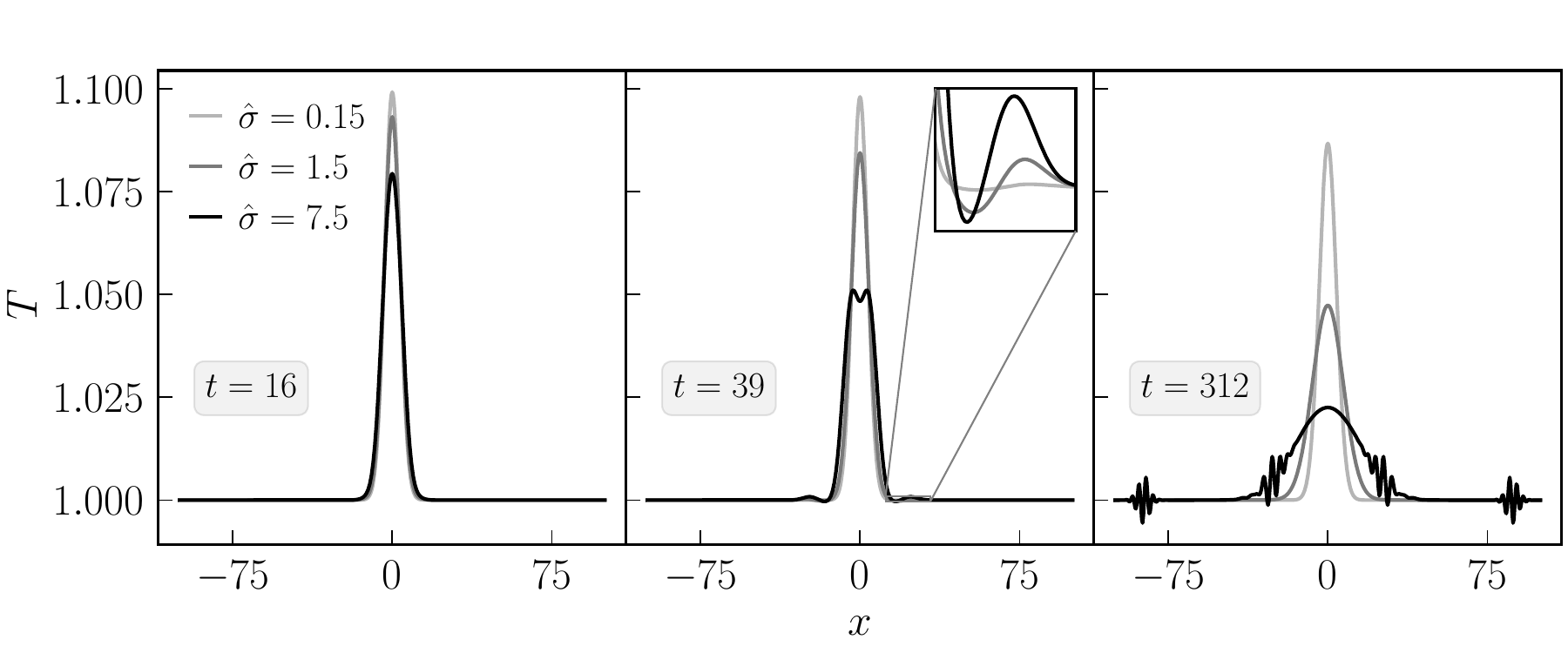}
\caption{Evolution of the heat flow initial data (\ref{eq:heat_flow_ID}) for three different BDNK fluids with $\hat{\sigma} = 0.15, 1.5, 7.5$, with the appropriate values of $\hat{\tau}$ such that $\hat{\sigma}/\hat{\tau} = 0.1$ in each case; other parameters are listed in Table \ref{table:parameters}.  At the earliest time (left panel) the central hot spot at $x = 0$ decays and spreads, similar to the behavior found in solutions to the heat equation.  Shortly later, though, the $\hat{\sigma} = 7.5$ case begins exhibiting wave-like behavior and the central peak splits in two, as expected from a telegrapher's-type equation in the limit $\tau_{\epsilon} \to \infty$ keeping $c_B^2 \propto \hat{\sigma}/\hat{\tau}$ constant, where it becomes a wave equation (cf. \cite{Most:2021rhr} Fig. 2). Also shown in this panel is a zoomed-in inset, showing that all cases show \textit{some} wavelike behavior in the form of a small transient which propagates away at the sound speed. Finally, at late times the $\hat{\sigma} = 7.5$ solution exhibits an oscillatory instability which eventually crashes the numerical simulation.} \label{fig:telegraphers}
\end{figure*}

In Fig.~\ref{fig:telegraphers}, we test the intuition derived from the
constant-coefficient heat flow problem studied in the previous subsection.
There, it is shown that for constant transport coefficients the BDNK
solution obeys a modified telegrapher's equation (\ref{eq:heat_t_BDNK}).
Note that (\ref{eq:heat_t_BDNK}) reduces to a simple 1D wave equation in
the limit $\sigma, \tau_{\epsilon} \to \infty$ so long as $c_B^2 \propto
\sigma/\tau_{\epsilon}$ is kept finite; in Fig. \ref{fig:telegraphers} we
numerically approximate this limit and observe the transition from
heat-equation-like behavior at small $\hat{\sigma}, \hat{\tau}$ (in light
gray) to wavelike behavior at large $\hat{\sigma}, \hat{\tau}$ (in black).
This wavelike behavior is readily apparent in the middle panel of Fig.
\ref{fig:telegraphers} for the $\hat{\sigma} = 7.5$ case, as the central
temperature peak splits in two rather than simply decaying.  That said, all
solutions (even the one with $\hat{\sigma} = 0.15$) possess \textit{some}
wavelike behavior, which can be seen in the form of a transient shown in
the inset.

Though all three solutions shown in Fig. \ref{fig:telegraphers} possess
subluminal characteristic speeds, two of the three cases ($\hat{\sigma} =
1.5, 7.5$) violate the linear stability constraint
(\ref{eq:simple_constraints}) on $\hat{\sigma}$.  The effects of this
violation are readily apparent in the rightmost panel of Fig.
\ref{fig:telegraphers}, where the $\hat{\sigma} = 7.5$ solution exhibits an
oscillatory instability which eventually crashes the simulation.  
The $\hat{\sigma} = 1.5$ solution does not
appear to demonstrate any unexpected behavior, however; this may be because
the instability is slow relative to the dynamical timescale, or perhaps
that this case is somehow stabilized by a nonlinear mechanism.

\section{Conclusion} \label{sec:conclusion}

In this work we have presented the first relativistic fluid model with ideal gas microphysics based in the BDNK formalism, which rigorously proves causality, linear stability of equilibrium states, strong hyperbolicity, and consistency with the second law of thermodynamics, provided the transport coefficients obey two simple constraints (Sec. \ref{sec:model}).  This model serves as a significant extension over previous work \cite{Pandya_2021,Pandya:2022pif,Bantilan:2022ech} which has been restricted to conformal fluids at zero baryon chemical potential, and thus cannot capture the effects of bulk viscosity, ``nontrivial'' heat flow (see Footnote \ref{footnote:heat_flow}), or backreaction of the baryon current onto the stress-energy tensor.

Using this model, we investigated a number of properties of BDNK theory, though we restricted to four-dimensional flat spacetimes with high degrees of spatial symmetry.  We began with a comparison of the structure of the Eckart, MIS, and BDNK theories, and how each applies dissipation to the solution.  We pointed out that Eckart theory does so in an acausal fashion, applying dissipation instantaneously, whereas the MIS and BDNK theories apply dissipation through a ``relaxation''-type mechanism.  This mechanism arises in MIS theory by construction, as its dissipative degrees of freedom are defined to obey ``relaxation''-type PDEs.  In BDNK theory, it stems from the addition of terms which are higher order on-shell (in other words, proportional to projections of the relativistic Euler equations) to the stress-energy tensor.

Afterward we considered uniformly expanding, boost-invariant Bjorken flows for the BDNK fluid with ideal gas microphysics.  We investigated the limits of infinite and zero relaxation times, and find that the former leads to solutions which never equilibrate (as expected) and the latter leads to superluminal characteristics (also expected), though there is no clear qualitative impact on the solution in the latter case, other than that the equations become stiff and require small steps to integrate numerically.  This intuition was clarified in an investigation of planar shockwave solutions, where we find that solutions with superluminal characteristics show no indications of ill behavior (and/or acausal propagation) up to a certain threshold, beyond which increasing the characteristic speed results in the onset of a very fast instability.  We extended the intuition described in \cite{Pandya_2021,Freistuhler2021} regarding the existence of shockwave solutions for a conformal BDNK fluid to the general case, and provided evidence that shockwave solutions exist so long as the flow's three-velocity $v$ is never equal to the local characteristic speeds of the BDNK PDEs.

A model such as the one presented here is required to study ``pure'' heat flow solutions, where energy flows due to a thermal gradient while the pressure is kept constant.  These solutions do not exist in the conformal fluid model studied previously \cite{Pandya_2021,Pandya:2022pif} as there temperature gradients imply pressure gradients, since $P \propto T^{4}$.  In the context of the ideal gas model with constant transport coefficients, we investigated how the hydrodynamic frame impacts the causal properties of the solution, showing that heat flows according to the nonrelativistic heat equation in Eckart theory, a telegrapher's (damped wave) equation in a hybrid Eckart-BDNK frame, and finally a generalized telegrapher's equation in the BDNK class of frames.  We also provided an explanation for the thermodynamic instability of Eckart theory in this setup, and illustrate how BDNK theory manages to avoid this instability.  For non-constant coefficients, we posed a set of constant-pressure initial data and observe that nonzero thermal conductivity is indeed required for nontrivial heat flow solutions, and find that telegrapher's-equation-like behavior is indeed obtained for the proposed fluid model with frame (\ref{eq:hydro_frame}).  Finally, we studied what occurs when one violates the linear stability constraints (which reduce to an upper bound on the dimensionless thermal conductivity). For sufficiently small violations of the bound, there is no apparent qualitative effect, possibly owing to the fact that the bound only guarantees \textit{linear} stability, or perhaps implying that the onset of instability is slow compared to the dynamical timescale of the problem.  For sufficiently large violations of the bound, however, we observe the onset of an oscillatory instability.

We find in our investigation of Bjorken flow that one can obtain solutions which, at early times, have hydrodynamic parameters outside of the allowed physical ranges of their equilibrium counterparts---namely, we have temperatures $T < 0$.  This is allowed outside of equilibrium, as there the hydrodynamic variables are not unique, with the only restriction being that they relax to reasonable equilibrium values (i.e. $T > 0$).  It may cause problems when evaluating the equation of state, however, which assumes as inputs values of the \textit{equilibrium} hydrodynamic variables, and will not be defined for, e.g., $T < 0$.  We remark that this issue must be solved in order to apply realistic (tabulated) nuclear equations of state, but for the simple analytic one considered here, (\ref{eq:EOS}), it does not cause problems; we leave investigation of this issue to a future work.

As mentioned previously, the BDNK fluid with ideal gas microphysics has the potential to be of use in studies of pure hydrodynamics such as this one, as it is relatively simple, the microphysics are well-motivated, and the model contains all of the dissipative effects present in the BDNK formalism (shear viscosity, bulk viscosity, and heat conduction).  The gamma-law equation of state is also commonly used in toy models of astrophysical phenomena, examples of which include neutron stars \cite{Noble2015,Mosta2013} and black hole accretion disks \cite{Gammie2003,Stone2008}, both of which may be systems where dissipative fluid effects could be relevant at the level of next-generation observations \cite{Most2021bulk,Most:2022yhe,Chandra2017,EventHorizonTelescope:2022urf}.  Since the conformal fluid is contained as a limiting case of the ideal gas model, one may also use it as a somewhat generalized toy model for the quark-gluon plasma produced in heavy-ion collisions, though truly realistic studies will include a tabulated equation of state derived from nuclear theory.

Each of the systems mentioned above presents a number of avenues for future work.  Given that the ideal gas model includes bulk viscosity and heat conduction, one could perform a comparison with a MIS-type theory with the same microphysics in order to better understand the differences between the two formalisms.  BDNK theory has yet to be applied outside of flat spacetime, and the model proposed in this work would allow for investigations of viscous fluids in strong gravity (e.g. neutron stars and black hole accretion disks, as mentioned previously). That said, a truly ``modern'' study in most cases relevant to astrophysics will require a dissipative formulation of magnetohydrodynamics, examples of which exist but have yet to be applied in such contexts \cite{Armas:2022wvb,Most:2021uck,Most:2021rhr}.

\begin{acknowledgments}
This material is based upon work supported by the National Science
Foundation (NSF) Graduate Research Fellowship Program under Grant No.
DGE-1656466. Any opinions, findings, and conclusions or recommendations
expressed in this material are those of the authors and do not necessarily
reflect the views of the National Science Foundation. FP acknowledges
support from NSF Grant No.  PHY-2207286, the Simons Foundation, and the
Canadian Institute For Advanced Research (CIFAR). ERM acknowledges support
from postdoctoral fellowships at the Princeton Center for Theoretical
Science, the Princeton Gravity Initiative, and the Institute for Advanced
Study.
\end{acknowledgments}

\appendix

\section{Deriving suitable hydrodynamic frames} \label{sec:hydro_frame}
BDNK theory is proven to be causal, linearly stable about equilibrium, locally well-posed, strongly hyperbolic, and consistent with the second law of thermodynamics so long as the set of transport coefficients $\{\tau_{\epsilon}, \tau_{P}, \tau_{Q}, \eta, \zeta, \sigma, \beta_{\epsilon}, \beta_{n}\}$ satisfy a set of algebraic constraints.  In this section, we describe the process of deriving a choice for these transport coefficients (a ``hydrodynamic frame'') which simultaneously satisfies all of the aforementioned constraints.

In \cite{Bemfica:2020zjp} it is proven that BDNK theory is consistent with the second law of thermodynamics (in the sense that total entropy generation is nonnegative) up to third order in gradients provided one takes $\eta, \zeta, \sigma \geq 0$.  For the work that follows, we will assume the following about the parameters which appear:
\begin{equation} \label{eq:trans_coeff_ranges}
\begin{aligned}
&\rho, n, \tau_{\epsilon}, \tau_{P}, \tau_{Q}, \eta > 0 \\ 
&m, \zeta, \sigma \geq 0 \\
&0 < c_s^2 < 1, ~~~ 0 < \omega < 3 - 2 \sqrt{2} \approx 0.2, ~~~ 1 \leq \alpha,
\end{aligned}
\end{equation}
where $c_s^2, \omega, \alpha$ are defined in (\ref{eq:cs_sq}-\ref{eq:omega}).  Thus assuming (\ref{eq:trans_coeff_ranges}) guarantees consistency of our solutions with the second law of thermodynamics, up to third order in gradients.

Causality is established provided the transport coefficients satisfy equation (20) of \cite{Bemfica:2020zjp}, which we will reproduce here as:
\begin{align}
&\rho \tau_{Q} > \eta \tag{CAUS A} \label{eq:CAUS_A}\\
&B^2 \geq 4 A C \geq 0 \tag{CAUS B} \label{eq:CAUS_B}\\
&2 A > -B \geq 0 \tag{CAUS C} \label{eq:CAUS_C}\\
&A > -B - C \tag{CAUS D} \label{eq:CAUS_D}
\end{align}
where we have written the constraints in much simpler form using the shorthand of \cite{Bemfica:2020zjp},
\begin{align}
A &= \rho \tau_{\epsilon} \tau_{Q} \label{eq:A}\\
B &= - \tau_{\epsilon} \Big( \rho c_s^2 \tau_{Q} + V + \sigma \kappa_{s} \Big) - \rho \tau_{P} \tau_{Q} \label{eq:B}\\
C &= \tau_{P} (\rho c_s^2 \tau_{Q} + \sigma \kappa_{s}) - \beta_{\epsilon} V \label{eq:C}\\
D &\equiv \rho c_s^2 (\tau_{\epsilon} + \tau_{Q}) + V + \sigma \kappa_{\epsilon} \\
E &\equiv \sigma (p'_{\epsilon} \kappa_{s} - c_s^2 \kappa_{\epsilon}).
\end{align}
In terms of this shorthand, the (squared) characteristic speeds $c_{\pm}^{2}, c_{1}^2$ are given by (\ref{eq:cpmsq}-\ref{eq:c1sq}).

Local well-posedness and strong hyperbolicity are established given $\eta, \zeta, \sigma \geq 0$, (\ref{eq:CAUS_A}-\ref{eq:CAUS_D}), and some restrictions on the allowed initial data.  Linear stability of solutions about thermodynamic equilibrium is given provided one satisfies equation (48) of \cite{Bemfica:2020zjp}, reproduced here:
\begin{align}
&(\tau_{\epsilon} + \tau_{Q}) |B| \geq \tau_{\epsilon} \tau_{Q} D \tag{STAB A1} \label{eq:STAB_A1}\\
&\tau_{\epsilon} \tau_{Q} D \geq \rho c_s^2 \tau_{\epsilon} \tau_{Q} (\tau_{\epsilon} + \tau_{Q}) \tag{STAB A2} \label{eq:STAB_A2} 
\end{align}
\begin{multline}
(\tau_{\epsilon} + \tau_{Q}) |B| D + \rho \tau_{\epsilon} \tau_{Q} (\tau_{\epsilon} + \tau_{Q}) E > \\
\tau_{\epsilon} \tau_{Q} D^2 + \rho (\tau_{\epsilon} + \tau_{Q})^2 C \tag{STAB B} \label{eq:STAB_B}
\end{multline}
\begin{equation}
c_s^2 D - E \geq \rho c_s^4 (\tau_{\epsilon} + \tau_{Q}) \tag{STAB C} \label{eq:STAB_C}
\end{equation}
\begin{multline}
(\tau_{\epsilon} + \tau_{Q}) \Big[ |B| (c_s^2 D - 2 E) + 2 c_s^2 \rho \tau_{\epsilon} \tau_{Q} E + C D \Big] > \\
2 c_s^2 \rho (\tau_{\epsilon} + \tau_{Q})^2 C + \tau_{\epsilon} \tau_{Q} D (c_s^2 D - E) \tag{STAB D} \label{eq:STAB_D}
\end{multline}
\begin{multline}
|B| D [C (\tau_{\epsilon} + \tau_{Q}) + E \tau_{\epsilon} \tau_{Q}] + 2 \rho \tau_{\epsilon} \tau_{Q} (\tau_{\epsilon} + \tau_{Q}) C E > \\
\rho C^2 (\tau_{\epsilon} + \tau_{Q})^2 + \tau_{\epsilon} \tau_{Q} (C D^2 + \rho \tau_{\epsilon} \tau_{Q} E^2) + B^2 E (\tau_{\epsilon} + \tau_{Q}), \tag{STAB E} \label{eq:STAB_E}
\end{multline}
where we have split their (48a) into our (\ref{eq:STAB_A1}-\ref{eq:STAB_A2}).  In summary, the hydrodynamic frames we are interested in here simultaneously satisfy (\ref{eq:trans_coeff_ranges}), (\ref{eq:CAUS_A}-\ref{eq:CAUS_D}), and (\ref{eq:STAB_A1}-\ref{eq:STAB_E}).  Of these three sets of inequalities, it is clear from inspection that the linear stability constraints (\ref{eq:STAB_A1}-\ref{eq:STAB_E}) are the most complicated, so we will focus on satisfying those first.

We work through the following two subsections in a publicly available Mathematica notebook\footnote{\url{https://github.com/aapandy2/BDNK_frame_constraints}}, which allows for analytic simplification of the complicated inequalities above.  We provide the notebook so that others may confirm the calculations done here, and so that they may modify the notebook to evaluate the BDNK constraints for other classes of hydrodynamic frames.

\subsection{Linear stability constraints}

Inspection of (\ref{eq:STAB_A1}-\ref{eq:STAB_E}) shows that the inequalities contain both the shorthand $B, C, D, E$ as well as explicit factors of $\rho, c_s^2, \tau_{\epsilon}, \tau_{Q}$.  We will begin by rescaling $B, C, D, E$ to absorb these factors to get:
\begin{equation} \label{eq:rescaled_shorthand}
\begin{aligned}
\hat{B} &\equiv \frac{B}{\rho c_s^2 \tau_{\epsilon} \tau_{Q}} = - \Big[ 1 + \frac{L \hat{V}}{\tau_{Q}} (1 - \omega \hat{\sigma}) + \frac{\tau_{P}}{c_s^2 \tau_{\epsilon}} \Big] \\
\hat{C} &\equiv \frac{C}{\rho c_s^4 \tau_{\epsilon} \tau_{Q}} = \frac{\tau_{P}}{c_s^2 \tau_{\epsilon}} + \frac{L \hat{V}}{\tau_{\epsilon}} \Big( \frac{L \hat{V}}{\tau_{Q}} \hat{\sigma} - \frac{\tau_{P}}{c_s^2 \tau_{Q}} \hat{\sigma} \omega - \alpha \Big) \\
\hat{D} &\equiv \frac{D}{\rho c_s^2 (\tau_{\epsilon} + \tau_{Q})} = 1 + \frac{L \hat{V}}{(\tau_{\epsilon} + \tau_{Q})} (1 - \hat{\sigma}) \\
\hat{E} &\equiv \frac{E}{\rho c_s^4 (\tau_{\epsilon} + \tau_{Q})} = \frac{\hat{\sigma} L \hat{V}}{(\tau_{\epsilon} + \tau_{Q})} (1 - \alpha \omega) \\
\end{aligned}
\end{equation}
where we have further defined
\begin{equation}
\begin{aligned}
V &= \hat{V} \, L \rho c_s^2 \\
\sigma &= \hat{\sigma} \, \frac{L \hat{V} \rho c_s^2}{(-\kappa_{\epsilon})}
\end{aligned}
\end{equation}
and we have used (\ref{eq:omega}-\ref{eq:alpha}).  In terms of $\hat{B}, \hat{C}, \hat{D}, \hat{E}$ the linear stability constraints (\ref{eq:STAB_A1}-\ref{eq:STAB_E}) become
\begin{equation} \label{eq:rescaled_constraints}
\begin{aligned}
&|\hat{B}| \geq \hat{D} \\
&\hat{D} \geq 1 \\
&|\hat{B}| \hat{D} +  \hat{E} - \hat{D}^2 - \hat{C} > 0 \\
&\hat{D} - \hat{E} \geq 1 \\
&|\hat{B}| \hat{D} + \hat{E} - \hat{D}^2 - \hat{C} > 2 |\hat{B}| \hat{E} + \hat{C} - \hat{D} \hat{E} - \hat{E} - \hat{C} \hat{D} \\
&\Big[ |\hat{B}| \hat{D} + \hat{E} - \hat{D}^2 - \hat{C} \Big] \hat{C} > 
\Big[ \hat{E} + \hat{B}^2 - \hat{C} - |\hat{B}| \hat{D} \Big] \hat{E},
\end{aligned}
\end{equation}
which appear to be much simpler.  In particular, the first, second, and fourth lines are especially simple; substituting the definitions (\ref{eq:rescaled_shorthand}) into these three constraints and using the parameter ranges (\ref{eq:trans_coeff_ranges}) one finds that the first line is automatically satisfied, and it is clear that since $\hat{E} \geq 0$ (by thermodynamics; see \cite{Bemfica:2020zjp}) the fourth line implies the second one.  Thus the three ``simple'' linear stability constraints boil down to just the fourth line of (\ref{eq:rescaled_constraints}), which is
\begin{equation} \label{eq:simple_stab_const}
1 - (2 - \alpha \omega) \hat{\sigma} \geq 0;
\end{equation}
noting that $0 < \alpha \omega \lessapprox 0.2$ (\ref{eq:trans_coeff_ranges}) we can see that the strongest case of (\ref{eq:simple_stab_const}) has $\alpha \omega = 0$ so (\ref{eq:simple_stab_const}) is implied by $1 - 2 \hat{\sigma} \geq 0$, or equivalently $\hat{\sigma} \leq \frac{1}{2}$.

Moving on to the three ``complicated'' linear stability constraints, we find it essential to make use of a computer algebra system such as Mathematica\footnote{When using such a system, it is necessary to minimize the number of free parameters appearing in the constraints; we find that computation time is significantly reduced in the rescaled quantities (\ref{eq:rescaled_shorthand}) in that the only free parameters which appear are $\alpha, \omega, \hat{\tau}, \hat{V}, \hat{\sigma}$.} along with an ansatz for the relaxation times $\tau_{\epsilon}, \tau_{P}, \tau_{Q}$.  We find the class of frames of the form
\begin{equation} \label{eq:frame_ansatz}
\tau_{\epsilon} = \tau_{Q} = L \hat{V} \, \hat{\tau}, ~~~ \tau_{P} = 2 \alpha c_s^2 L \hat{V}
\end{equation}
where we require $\hat{\tau}, L > 0$ for consistency with (\ref{eq:trans_coeff_ranges}), along with the further restriction
\begin{equation} \label{eq:sigma_bound}
\hat{\sigma} \leq \frac{1}{3}
\end{equation}
satisfies all of the linear stability inequalities, (\ref{eq:STAB_A1}-\ref{eq:STAB_E}).

\subsection{Causality constraints}
Given the assumptions (\ref{eq:trans_coeff_ranges}), the frame ansatz (\ref{eq:frame_ansatz}), and the constraint (\ref{eq:sigma_bound}), one can show that (\ref{eq:CAUS_B}) and the second half of (\ref{eq:CAUS_C}) are automatically satisfied, leaving just (\ref{eq:CAUS_A}), the first half of (\ref{eq:CAUS_C}), and (\ref{eq:CAUS_D}).  These three remaining constraints take the form
\begin{equation} \label{eq:caus_const_simplified}
\begin{aligned}
\hat{\tau} &> c_s^2 \, \frac{\eta}{\frac{4 \eta}{3} + \zeta} \\
2 \hat{\tau} &> c_s^2 (2 \alpha - \omega \hat{\sigma} + \hat{\tau} + 1) \\
c_s^4 (-2 \alpha \omega \hat{\sigma} + \hat{\sigma} + \alpha \hat{\tau}) + \hat{\tau}^2 &\geq c_s^2 \hat{\tau} (2 \alpha - \omega \hat{\sigma} + \hat{\tau} + 1).
\end{aligned}
\end{equation}
All three of these inequalities are implied by the stronger, single inequality
\begin{equation} \label{eq:fully_simplified_caus_const}
\hat{\tau} \geq \frac{(\Gamma - 1)(2 - c_s^2) + c_s^2}{1 - c_s^2},
\end{equation}
which is obtained from the third line of (\ref{eq:caus_const_simplified}) in the limit ${\sigma \to 0}$, which increases the lower bound on $\hat{\tau}$.  The tradeoff for using (\ref{eq:fully_simplified_caus_const}) rather than (\ref{eq:caus_const_simplified}) is that saturating (\ref{eq:fully_simplified_caus_const}) for $\hat{\sigma} > 0$ will yield a maximum characteristic speed strictly smaller than the speed of light; saturating (\ref{eq:caus_const_simplified}), on the other hand, always gives $|c_+| = 1$.  Note that \cite{Bemfica:2020zjp} excludes the case $|c_+| = 1$ from their proofs; if one wishes to ensure that the hydrodynamic frame has $|c_+| < 1$, one must alter the third line of (\ref{eq:caus_const_simplified}) and (\ref{eq:fully_simplified_caus_const}) to have $>$ rather than $\geq$.

For the frame used in this study (\ref{eq:hydro_frame}), the (squared) characteristic speeds are
\begin{multline} \label{eq:cpmsq}
c_{\pm}^{2} = \frac{c_s^2}{2 \hat{\tau}} \Big(2 \alpha -\omega  \hat{\sigma} + \hat{\tau} + 1 \pm \Big[ \omega \hat{\sigma} (4 \alpha +\omega  \hat{\sigma})+(2 \alpha +1)^2 \\
- 2 (\omega +2) \hat{\sigma}+\hat{\tau}^2+\hat{\tau} (2-2 \omega  \hat{\sigma})\Big]^{1/2} \Big)
\end{multline}
\begin{equation} \label{eq:c1sq}
c_{1}^{2} = c_s^2 \frac{\eta}{V \hat{\tau}},
\end{equation}
where $V$ is defined in (\ref{eq:V}).

\section{Numerical algorithms and convergence tests} \label{sec:numerics}
Throughout this work we integrate all ODEs---namely (\ref{eq:Bjorken_EOM}) in Sec. \ref{sec:Bjorken_flow} and (\ref{eq:shockwave_nprime}, \ref{eq:shockwave_epsP}-\ref{eq:shockwave_velP}) in Sec. \ref{sec:shockwaves}---using the fourth-order explicit Runge-Kutta method, RK4.  These solutions are produced at resolutions ranging between $N = 2^{9}$ and $2^{13}$ gridpoints, and the rate of convergence
\begin{equation} \label{eq:convergence_factor}
Q_{N} = \frac{||R_{N/2}||}{||R_{N}||},
\end{equation}
is computed and summarized in Table \ref{table:ODE_conv}.  In (\ref{eq:convergence_factor}), $R_{N}$ is a discrete residual evaluated on the numerical solution of resolution $N$ (whose value should be identically zero for solutions to the continuum PDEs) and $||\cdot||$ is any vector norm (here the 1-norm).  One may straightforwardly show using the Richardson expansion that a numerical scheme that is second order in the grid spacing will have $Q_{N} \to 4$ as $N \to \infty$, and a fourth-order scheme should have $Q_{N} \to 16$.  All ODE solutions presented in this work demonstrate fourth-order convergence as the grid is refined, consistent with the use of the RK4 method.

\begin{table}[h]
\renewcommand{\arraystretch}{1.2}
\begin{tabular}{c || c | c | c | c}
Test & N & $Q_{N/4}$ & $Q_{N/2}$ & $Q_{N}$ \\ \hline \hline
Bjorken flow, $\hat{\tau} = 0.5$ & $2^{11}$ & 34.8 & 18.7 & 16.9 \\ 
Bjorken flow, $\hat{\tau} = 1$ & $2^{11}$   & 18.4 & 16.9 & 16.3   \\
Bjorken flow, $\hat{\tau} = 2$ & $2^{11}$   & 16.9 & 16.3 & 16.1   \\
Shockwave                      & $2^{13}$   & 15.9 & 15.9 & 15.9  \\
\end{tabular}
\caption{Convergence factor $Q_{N}$ (\ref{eq:convergence_factor}), for three resolutions $N/4, N/2, N$, where $N$ is given in the second column of the table, for the Bjorken flow solutions with $\dot{\epsilon} = -2$ (convergence results for solutions with $\dot{\epsilon} = 0, 2$ are essentially identical) and the shockwave ODE solution shown in Fig. \ref{fig:shockwave_profile}.  In all cases, we should observe $Q_{N} \to 16$ as $N \to \infty$.  For the Bjorken flow solutions, the residual $R_{N}$ is an independent (fourth-order centered finite difference) discretization of (\ref{eq:Bjorken_EOM}); for the shockwave case we use a fourth-order centered finite difference discretization of the $t$-component of (\ref{eq:Tab_cons_law}), namely $T^{tx}_{,x} = 0$.} \label{table:ODE_conv}
\end{table}

\begin{figure}[h]
\centering
\includegraphics[width=1\columnwidth]{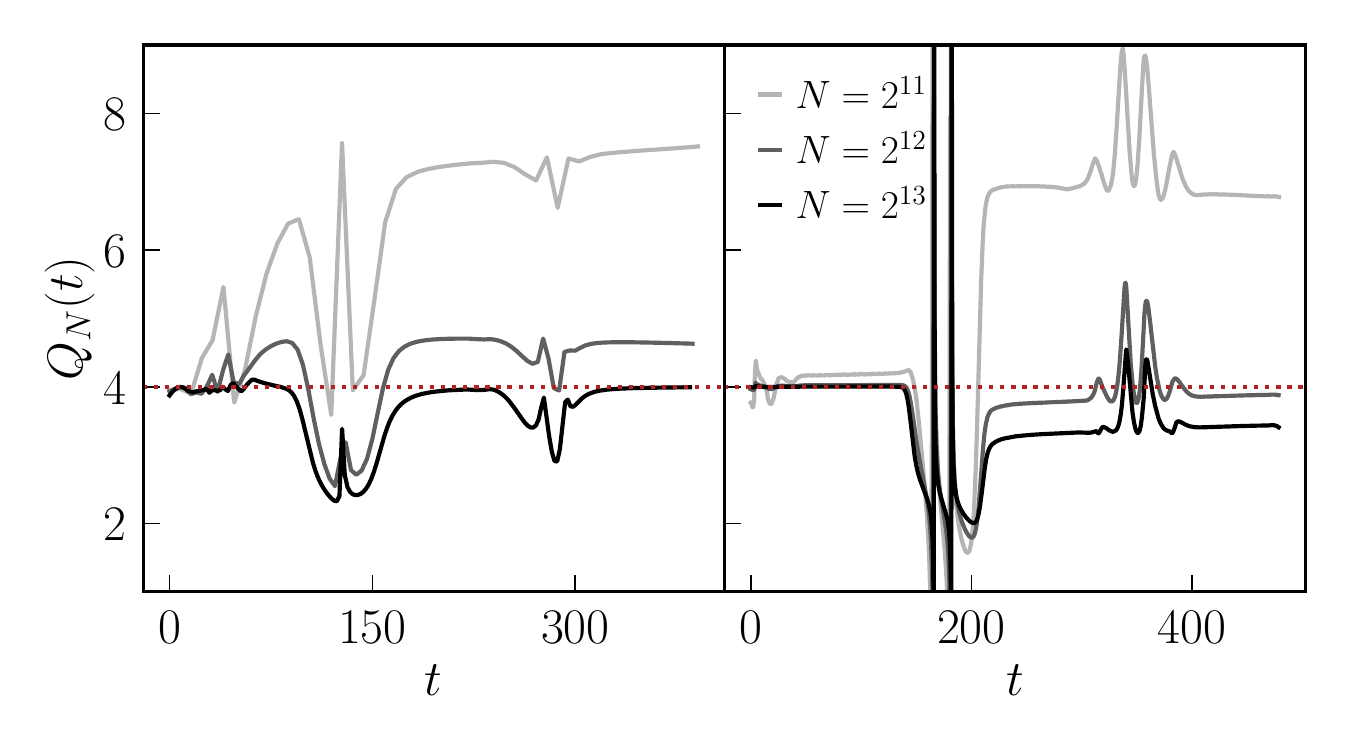}
\caption{Convergence plot showing $Q_{N}(t)$, (\ref{eq:convergence_factor}), computed with an independent (second-order Crank-Nicolson) discretization of the $t$-component of (\ref{eq:Tab_cons_law}) for the stable shockwave solution shown in the bottom panel of Fig. \ref{fig:shock_instability} as well as the $\hat{\sigma} = 0.15$ case of the heat flow problem shown in Fig. \ref{fig:telegraphers}.  The numerical scheme is second-order so we expect to find $Q_{N} \to 4$ as $N \to \infty$, which occurs in each case near $t = 0$ up until there is a significant interaction with the boundary due to transients from the initial data propagating away ($t \sim 80$ in the left panel, $t \sim 150$ in the right panel).  Afterward, the solutions converge at a rate between first and second order.} \label{fig:conv_plot}
\end{figure}

To solve the BDNK PDEs, we employ the conservative finite volume method of \cite{Pandya:2022pif}.  It uses the method of lines to integrate the system forward in time with the total variation diminishing second-order Runge-Kutta scheme known as Heun's method.  We use CFL number $\lambda \equiv \frac{\Delta t}{\Delta x} = 0.1$ throughout, except for the ``stiff superluminal'' and ``wildly superluminal'' solutions in Fig. \ref{fig:acaus_instab} which require $\lambda = 0.01$.  In space, the method of \cite{Pandya:2022pif} uses a WENO/CWENO discretization which is at most fourth-order convergent in the grid spacing for smooth flows; thus, the scheme is second-order overall, but can converge at higher rates at finite resolution when time derivatives are small.  Convergence results are shown for the shockwave (Sec. \ref{sec:shockwaves}) and heat flow (Sec. \ref{sec:heat_flow}) problems in Fig. \ref{fig:conv_plot}.  In both cases, solutions converge at second order in the grid spacing, as expected, up until there is significant interaction with the ghost cell boundaries; afterward, the solution converges at a rate between first and second order.

\bibliography{references}

\end{document}